\documentclass[12pt]{iopart}
\usepackage{epsfig}
\usepackage{epstopdf}
\usepackage [latin1]{inputenc}
\usepackage{color}
\usepackage{enumerate}
\usepackage{dsfont}
\usepackage{cite}
\usepackage{mathrsfs}
\newcommand{\ket}[1]{{\left | {#1} \right\rangle}}
\newcommand{\bra}[1]{{\left\langle {#1} \right |}}
\newcommand{\av}[1]{{\left\langle {#1} \right\rangle}}

\providecommand{\keywords}[1]
{ \small  \textbf{\textit{Keywords---}} #1}

\begin{document}

\title{\bf 
Quantum information processing with superconducting circuits: a perspective}
\author{G. Wendin}
\address{Department of Microtechnology and Nanoscience - MC2, \\
Chalmers University of Technology, \\SE-41296 Gothenburg, Sweden}

\date{\today}

\begin{abstract}

\noindent

The last five years have seen a dramatic evolution of platforms for quantum computing, taking the field from physics experiments to quantum hardware and software engineering. Nevertheless, despite this progress of quantum processors, the field is still in the noisy intermediate-scale quantum (NISQ) regime, seriously limiting the performance of software applications. Key issues involve how to achieve quantum advantage in useful applications for quantum optimization and materials science, connected to the concept of quantum supremacy first demonstrated by Google in 2019. In this article we will describe recent work to establish relevant benchmarks  for quantum supremacy and quantum advantage, present recent work on applications of variational quantum algorithms for optimization and electronic structure determination, discuss how to achieve practical quantum advantage, and finally outline current work and ideas about how to scale up to competitive quantum systems.

\end{abstract}


\maketitle

\tableofcontents

\newpage

%%%%%%%%%%%%%%%%%%%%%%%%%%%%%%%%%%%%%%%%%%%
%%%%%%%%%%%%%%%%%%%%%%%%%%%%%%%%%%%%%%%%%%%
%%%%%%%%%%%%%%%%%%%%%%%%%%%%%%%%%%%%%%%%%%%
%%%%%%%%%%%%%%%%%%%%%%%%%%%%%%%%%%%%%%%%%%%

\section{Introduction}

Since around 1980, quantum computing (QC) and quantum simulation (QS) have gone from fantasy to possibility, from concept to application, from basic science to engineering \cite{Wendin2017,Gu2017,Krantz2019,Kjaergaard2020,Blais2020,Blais2021,Bruzewicz2019,Brown2021,Daley2022}.
In 2012, at a workshop at Benasque in the Spanish Pyrenees, at a memorable session we discussed in particular the future of QC. Myself, I predicted 20-30 years for useful applications, while Rainer Blatt emphasized that if we did not have any decisive results in 5 years, QC would soon be dead. It seems we were both right: QC did take off around 2017 at engineering levels, while really useful competitive applications showing practical quantum advantage are probably still 10-20 years ahead. The intense discussions in the European quantum community then led to the 2016 Quantum Manifesto \cite{QuantumManifesto2016} and to the EU Quantum Flagship \cite{QFlag} setting sail in 2018.

From 2019, the field has seen a huge development of platforms for quantum computing and simulation  with superconducting devices and systems \cite{Arute2019a,Wu2021,Zhu2022,Corcoles2020,Gambetta2022}, including demonstration of quantum supremacy \cite{Preskill2012,Arute2019a,Wu2021,Zhu2022}. However, we are living in the era of noisy intermediate-scale quantum (NISQ) devices  \cite{Preskill2018}, and it is currently impossible to build superconducting quantum processing units (QPU) where one can entangle more than about 20 qubits with high probability during the coherence time.  Which means there is no time for useful computation with deep quantum circuits, only time to characterize the device and demonstrate physical entanglement - impressive but  not necessarily useful. Nevertheless, IBM is now scaling superconducting QPUs to more than 1000 qubits in 2023 and over 4000 qubits in 2025 \cite {Gambetta2022,Bravyi2022}, aiming for seamless integration of high-performance  computers (HPC) and QPU accelerators. Google \cite{Lucero2021} seems to focus on modest-size quantum error corrected QPUs for large-scale quantum computational breakthroughs already by 2029, while at the same time there seems to be consensus \cite{Sanders2020,Babbush2021,Beverland2022} that practical quantum advantage may take much longer to achieve. 

To create useful applications showing quantum advantage, it is necessary to scale up QPUs and related classical-quantum hybrid (HPC+QC)  infrastructure.
This explains a number of current trends: (i) stay at "small" scales ($\le$ 100 qubits) and try to solve coherence problems and create useful applications before scaling up;  (ii) go for large scales ($\ge$ 1000 qubits) and try to implement quantum error correction for quantum advantage or superiority while scaling up; (iii) scale up and solve large-scale hardware (HW) and software (SW) integration at systems levels, waiting for practical quantum advantage for use cases to emerge.

The question of the feasibility of powerful quantum computers beating classical super-HPC hinges on that it will be ultimately possible to perform quantum error correction (QEC). When John Martinis' group was able to demonstrate that their superconducting quantum circuits were at the surface code threshold for fault tolerance  \cite{Barends2014}, then the field opened up and went from quantum physics toward quantum computing, scaling up HW and SW \cite{Barends2015,Barends2016,Song2017,Kandala2017,Kandala2019,Gong2019,Arute2019a,Gong2021,Gong2022},
and implementing significant quantum algorithms and quantum physics experiments
\cite{Barends2015,Barends2016,Kandala2017,Kandala2019,Gong2019,Arute2020a,Arute2020b,Neill2021,Gong2021,Gong2022,PZhang2022,Mi2022,Stanisic2022}, including demonstrations of significant steps toward QEC \cite{Andersen2020,Chen2021,Marques2022,Krinner2022,Acharya2022}.   

Much of the current discussion concerns how to get from proofs of concept to useful applications.  Consider the way quantum computing is being promoted by Google  \cite {Lucero2021}: "Within the decade, Google aims to build a useful, error-corrected quantum computer. This will accelerate solutions for some of the world's most pressing problems, like sustainable energy and reduced emissions to feed the world's growing population, and unlocking new scientific discoveries, like more helpful AI."

This tells us that the world's most pressing problems, like benchmarking climate models, are already subject to large-scale calculations, pushing super-HPCs to their limits set by NP-hard problems. 
The intended role of QPUs is to provide quantum superiority and to go far beyond those limits.
However, in the short term, during this decade, this can only be achieved by experimental quantum co-processors running specific subroutines  addressing classically hard problems, omnipresent in industrial use cases \cite{QUTAC2021,Kim2022}. In the short term, industry will effectively be co-developing quantum algorithms as subroutines, and benchmark them against competing classical algorithms.  If this leads to quantum advantage already in the short term, that will be a great bonus. The important thing to understand is that lack of quantum advantage for now does not jeopardize powerful computing - exascale super-HPC platforms will continue addressing the world's most pressing problems, and eventually QPU accelerators may provide quantum leaps.

This perspective article can be looked upon as a self-contained  second part of a research and review paper  \cite{Wendin2017} that stopped short at the beginning of the current engineering era of scaling up devices and building quantum computing ecosystems.  The purpose is to focus on addressing the quite dramatic development during the subsequent five years, trying to "predict the future" based on current visions, roadmaps, efforts and investments that aim for the next ten years \cite {Alexeev2021,Altman2021,Ang2022},  outlining a sustainable quantum evolution that hopefully survives the quantum hype \cite{Ezratty2022,DasSarma2022}.  To be able to do so in this brief article, we will frequently refer to \cite{Wendin2017} and to recent reviews for basic background, technology, and methods. The review will focus on superconducting technology and systems based on circuit quantum electrodynamics (cQED) \cite{Gu2017,Blais2021}, but will also provide glimpses of the broader development.

%%%%%%%%%%%%%%%%%%%%%%%%%%%%%%%%%%%%%%%%%%%
%%%%%%%%%%%%%%%%%%%%%%%%%%%%%%%%%%%%%%%%%%%
%%%%%%%%%%%%%%%%%%%%%%%%%%%%%%%%%%%%%%%%%%%
%%%%%%%%%%%%%%%%%%%%%%%%%%%%%%%%%%%%%%%%%%%

\section{Overview}

%%%%%%%%%%%%%%%%%%%%%%%%%%%%%%%%%%%%%%%%%%%
%%%%%%%%%%%%%%%%%%%%%%%%%%%%%%%%%%%%%%%%%%%
%%%%%%%%%%%%%%%%%%%%%%%%%%%%%%%%%%%%%%%%%%%

\subsection{Quantum processor systems: hardware and software}

\begin{figure}
\centering
\includegraphics[width=15cm]{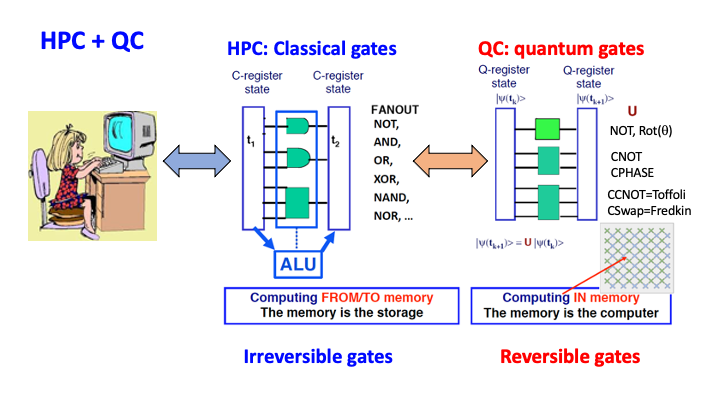} 
\caption{HPC + QC. The user prepares a program and submits it to the classical frontend. The HPC prepares the quantum circuit and sends it to the QC. The HPC/QC registers have N bits/qubits, i.e. $n=2^N$ possible configurations/states. The HPC register can only be in one of $2^N$ states: $\ket{00...00}, \ket{00...01}, \ket{00...10}, .....\ket{11..11} $ at each instance of time $t$, while the QC register can be in a  {\em superposition of all states}:  $ f_1(t) \ket{00...00} + f_2(t) \ket{00...01}  + f_3(t) \ket{00...10}, .....  + f_{n-1} (t)\ket{11...10}  + f_n(t) \ket{11...11} $. This describes {\em time-dependent quantum superposition and entanglement} and can, at best, lead to exponential quantum advantage.}
\label{Little_girl}
\end{figure}

With NISQ devices with limited coherence time, the necessary circuit depths \cite{circuitdepth} are much too long to achieve reasonable accuracy.
It is therefore necessary to break up the quantum circuits in short, low depth, pieces that can be run on quantum processing units (QPU) during the coherence time.
These often make use of variational quantum algorithms (VQA) where one calculates the expectation value of a cost function, e.g. a Hamiltonian, using a parameterized trial function.
A classical high-performance computer (HPC) controls and executes the classical optimization loop:  computes averages, searches for improved energies, computes new parameters, and updates the quantum circuit.
 
Figure~\ref{Little_girl}  illustrates the basics of quantum computing from a user perspective. 
The program code is typically prepared on a small classical computer, then submitted to the application programming interface (API) of an HPC frontend. The HPC interprets the quantum part and prepares the code defining the quantum circuit. This is finally loaded into the stack of the QPU. In the case of an ideal QPU, the code is executed on the QPU until solution is achieved, and the results finally read out and sent back to the HPC for post-processing.  In the NISQ world of QPUs, the quantum execution has to be limited to low-depth (shallow) quantum circuits that can be executed within the coherence time.  This necessitates repeated quantum-classical processing loops for optimization of variational problems. 

Here the classical computer is a bottleneck. The HPC calculation takes much longer time than the QPU execution time, even if the HPC responds without delay (no latency) and that the QPU backend is available immediately on request. This is hybrid HPC+QC computation, and is algorithm dependent. 

Quantum advantage in the NISQ era depends critically on efficient representation and coding of problems.
Here there is a distinct difference between decision and optimization problems on the one hand and, e.g., computational problems like electron structure and energy level determination on the other.

%%%%%%%%%%%%%%%%%%%%%%%%%%%%%%%%%%%%%%%%%%%
%%%%%%%%%%%%%%%%%%%%%%%%%%%%%%%%%%%%%%%%%%%
%%%%%%%%%%%%%%%%%%%%%%%%%%%%%%%%%%%%%%%%%%%

\subsection{Quantum algorithms}

An ideal digital quantum computer executes perfect gates on ideal qubits with infinite coherence time.
It evolves the time-evolution operator $e^{-iHt}$ corresponding to a given Hamiltonian describing the problem (see e.g. \cite{Wendin2017}). $e^{-iHt}$  is then broken down into a product of factors for the different terms of the Hamiltonian. These factors are finally represented in terms of quantum gates constituting a quantum circuit.
In this case, the role of a classical computer is exclusively pre- and post-processing: preprocessing to construct the quantum circuit and post-processing to read out and treat the results. Both of these are in principle NP-hard. To get desired results, the QC must be able to run for long times to execute deep quantum circuits, which requires perfect qubits and gates. With NISQ devices, it is not possible to run e.g. phase-estimation algorithms to compute the energies of molecules - the needed quantum circuits are far too long with respect to the coherence time. This has led to alternative approaches, calculating the expectation value of the problem Hamiltonian with respect  to parametrized trial functions and then optimizing the parameters for lowest energy.

Variational quantum algorithms (VQA) are generally based on constructing parametrized trial functions to compute and minimize the expectation value of a cost function. In the quantum case, the specific quantum computation involves computing the expectation value of a Hamiltonian cost function, while the classical computer prepares the trial function, computes the energy, updates the trial function parameters and minimizes the energy in an optimization loop.  Extensive discussions and reviews of quantum methods and algorithms are presented in \cite{Wang2018,Cerezo2021,Franca2021,Tilly2022,Bharti2022,Abhijith2022,Kowalsky2022}.

%%%%%%%%%%%%%%%%%%%%%%%%%%%%%%%%%%%%%%%%%%%
%%%%%%%%%%%%%%%%%%%%%%%%%%%%%%%%%%%%%%%%%%%
%%%%%%%%%%%%%%%%%%%%%%%%%%%%%%%%%%%%%%%%%%%

\subsection{Quantum supremacy}

\begin{figure}
\centering
\includegraphics[width=16cm]{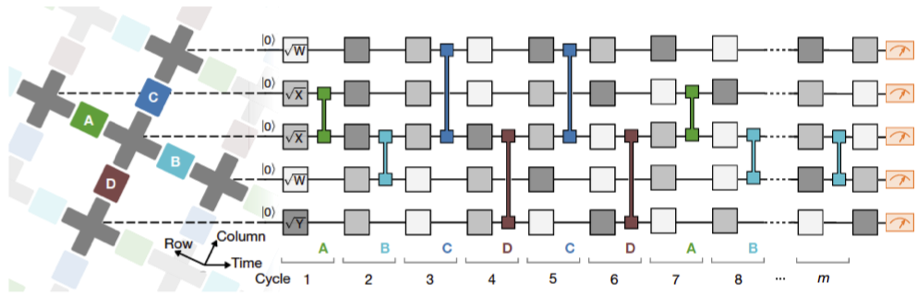} 
\caption{Control operations for generating the pseudo-random quantum circuits for Google's quantum supremacy bechmarking protocol \cite{Arute2019a}.  Adapted from \cite{Arute2019a}.}
\label{Google_QSup}
\end{figure}

John Preskill was the first one to explicitly introduce the concept of quantum supremacy in a 2012 paper discussing quantum computing and the entanglement frontier \cite{Preskill2012}.  In 2016, Boixo et al. then wrote a paper on how to characterize quantum supremacy in near-term devices \cite{Boixo2018} preparing for the 2019 Google experiment to demonstrate quantum supremacy \cite{Arute2019a}. 
The idea was to measure the output of a pseudo-random quantum circuit (Fig.~\ref{Google_QSup}) to produce a distribution of samples, and to compute the cross-entropy describing the "overlap" between the quantum and classical distributions (see Hangleiter and Eisert \cite{Hangleiter2022} for a review of quantum random sampling).

Aaronson and Chen \cite{Aaronson2017}  put this on complexity-theoretic foundations. They noted, that for the sampling tasks, not only simulation but even verification might need classical exponential time. 
This made it advantageous to directly consider the  probability of observed bitstrings, rather than the distribution of sampled bitstrings. 
To this end, they showed \cite{Aaronson2017} that there's a natural average-case hardness assumption (Heavy Output Generation, HOG), which has nothing to do with sampling, yet implies that no polynomial-time classical algorithm can pass a statistical test that is passed by the outputs of the quantum sampling procedure. 
The Quantum Volume benchmark of IBM  (see Sect. 2.4.2) is based on HOG.

As mentioned, Google based their quantum supremacy demonstration \cite{Arute2019a} on sampling quantum and classical distributions, calculating the cross-entropy as described in \cite{Boixo2018}.  Cross-entropy benchmarking (XEB) has the advantage that it provides deeper insight than HOG, including measures of fidelity, and allows tracing of the development from small processors to devices that can only be simulated approximately. 

The Google paper  \cite{Arute2019a} stated that an HPC would take 10 000 years. That statement immediately met with a rebuttal on the IBM Research Blog by Perdnault et al.\cite{Perdnault2019}, explaining that an ideal simulation of the same task in a conservative, worst-case estimate could be performed on a classical system in 2.5 days and with far greater fidelity. Therefore the quantum supremacy threshold had not been met by Google using 53 qubits.

This was of course valid criticism, but effectively just delaying the inevitable. An experiment that more decisively passed the quantum supremacy threshold was soon announced by Chinese researchers \cite {Wu2021,Zhu2022} using the Zuchongzhi processor, closely following Google's recipes, to demonstrate distinct quantum computational advantage. 
In the most recent experiment  \cite {Zhu2022} using 60-qubit 24-cycle random circuit sampling, the state-of-the-art HPC classical simulation would have taken tens of thousands of years, while Zuchongzhi 2.1 only took about 4.2 h, thereby significantly enhancing the quantum computational advantage. 

As emphasized by Perdnault et al.\cite{Perdnault2019}, quantum supremacy is a threshold that does not automatically certify the quantum processor to be useful for running useful algorithms. However, it does benchmark the quality of the processor. The original Google experiment used a 53-qubit quantum processor that implements a large two-qubit gate quantum circuit of depth 20, with 430 two-qubit and 1,113 single-qubit gates, and with predicted total fidelity of  $F_{XEB}  = 0.2\% > 0$. 
The condition for quantum supremacy:  $F_{XEB}  > 0$, is based on statistics of creating an ensemble of a million runs of quantum circuits. The problem is that to solve e.g. a quantum chemistry problem based on 53 qubits, the depth of the quantum circuit would have to be in the range of a million. 
What is needed for useful problems challenging HPCs is to be able to run perfect quantum circuits with a number of 2-qubit gates much larger than the circuit width (number of qubits).  The name of the game is how to achieve  practical quantum advantage.

%%%%%%%%%%%%%%%%%%%%%%%%%%%%%%%%%%%%%%%%%%%
%%%%%%%%%%%%%%%%%%%%%%%%%%%%%%%%%%%%%%%%%%%
%%%%%%%%%%%%%%%%%%%%%%%%%%%%%%%%%%%%%%%%%%%

\subsection{Performance metrics}

%%%%%%%%%%%%%%%%%%%%%%%%%%%%%%%%%%%%%%%%%%%
%%%%%%%%%%%%%%%%%%%%%%%%%%%%%%%%%%%%%%%%%%%

\subsubsection{Cross entropy benchmarking - XEB}

The task is to sample the $2^N$ bitstring output of a pseudo-random quantum circuit (Fig.~\ref{Google_QSup}).
Cross-entropy benchmarking (XEB) compares the probability for observing a bitstring experimentally with the corresponding ideal probability computed via simulation on a classical computer. 

For a given circuit, one collects the measured bitstring sample $\{x_{i}\}$ and computes the
linear XEB fidelity \cite{Arute2019a}
\begin{equation}
\label{XEB}
F_{XEB}  = 2^N \av{P(x_i)}_i - 1
\end{equation}
where $N$ is the number of qubits,  $P(x_i)$ is the probability of the {\em experimental} bitstring $\{x_{i}\}$ computed for the {\em ideal quantum circuit,} and the average is over the observed bitstrings. 
$F_{XEB}$ is correlated with how often one samples high-probability bitstrings. If the distribution is uniform, then  
$\av{P(x_i)}_i  = 1/2^N$ and $F_{XEB}  = 0$. 
Values of $F_{XEB}$ between 0 and 1 correspond to the probability that no error has occurred while running the circuit. 
In the Google case  \cite{Arute2019a}, the computed values are very small,  $F_{XEB} \sim 10^{-3}$. This may represent proof of principle, but hardly provides any useful result. One needs to have $F_{XEB} \sim 1$ to be able to run algorithms, useful or not. 

To demonstrate quantum supremacy one must achieve a high enough $F_{XEB}$ for a circuit with sufficient width and depth such that the classical computing cost of $P(x_i)$ for the full circuit is intractable.
$P(x_i)$ must be calculated classically by simulating the ideal quantum circuit, which is formallly intractable in the region of quantum supremacy. 
Since at least 2016 it has been understood that Random Circuit Sampling (RCS), the task to sample the $2^N$ bitstring output of a pseudo-random quantum circuit, will not scale to arbitrarily many qubits without error-correction \cite{Aaronson2017}. Bouland et al. \cite{Bouland2019} provided strong complexity theoretic evidence of classical hardness of RCS, placing it on par with the best theoretical proposals for supremacy.
However, very recently Aharonov et al. \cite{Aharonov2022,Brubaker2023} produced a polynomial time classical algorithm for sampling from the output distribution of a noisy random quantum circuit. This gives strong evidence that, in the presence of a constant rate of noise per gate, random circuit sampling (RCS) cannot be the basis of a {\em scalable} experimental violation of the extended Church-Turing thesis. Noise kills entanglement and makes RCS classically tractable (provided the HPC has enough memory to do the calculation).

However, the algorithm does not directly address finite-size RCS-based quantum supremacy experiments  \cite{Aharonov2022}, so the result is not directly applicable to current attempts to invalidate the quantum supremacy results \cite{Arute2019a,Wu2021,Zhu2022} using classical HPC. Feng and Pan \cite{FengPan2022b} solved the Google sampling problem classically in about 15 h on a computational cluster with 512 GPUs with state fidelity 0.0037 (Google 0.00224), and claimed that it would only take few dozen seconds on an exascale machine, much faster than Google. 

Clearly it provides some satisfaction to demonstrate in practice that an HPC can beat the noisy 53q Sycamore QPU. However, a more challenging target for the HPC may now be to beat the 66q Zuchongzhi 2.1 with its  60-qubit 24-cycle RCS \cite{Zhu2022}.

%%%%%%%%%%%%%%%%%%%%%%%%%%%%%%%%%%%%
%%%%%%%%%%%%%%%%%%%%%%%%%%%%%%%%%%%%

\subsubsection{Quantum volume - QV} 
\begin{figure}
\centering
\includegraphics[width=10cm]{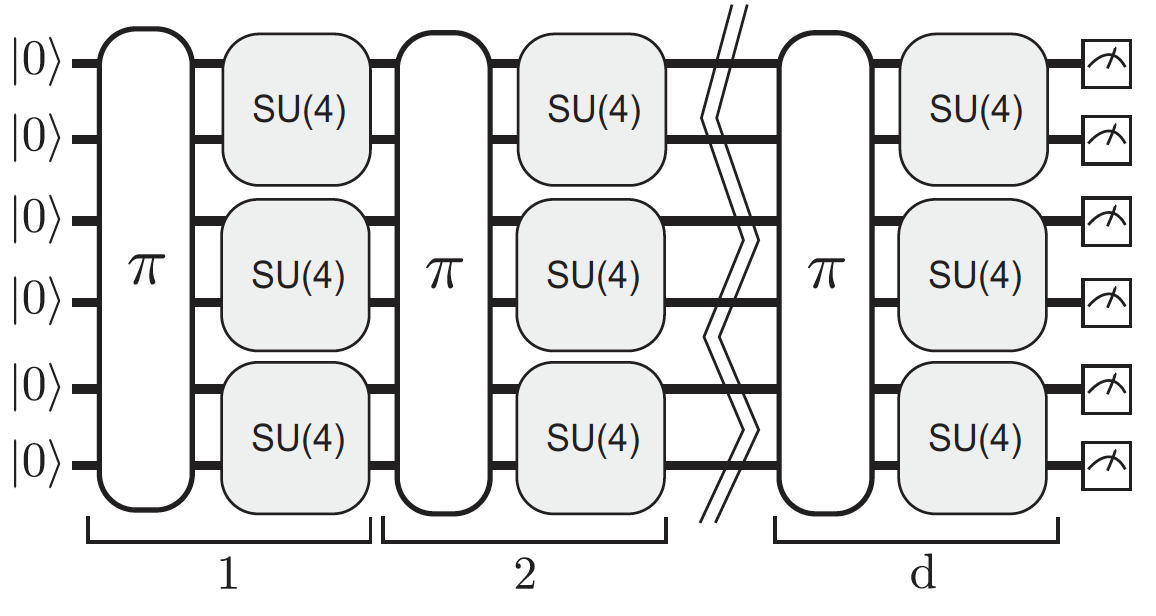} 
\caption{IBM QV pseudo-random quantum circuit \cite{Cross2019} consisting of $d$ layers (depth) of random permutations $\pi$ of the $N$ qubit labels, followed by random SU(4) two-qubit
gates. When the circuit width $N$ is odd, one of the qubits is idle in each layer. From  \cite{Cross2019}.}
\label{QV_PRQC}
\end{figure}

The fundamental challenges in the NISQ era can be illustrated using the concept of Quantum Volume (QV) introduced by IBM \cite{Cross2019}.   QV is linked to system error rates, and quantifies the largest random circuit of equal width and depth that a specific computer can successfully implement given decoherence, gate fidelities, connectivity, and more \cite{Cross2019,Pelofske2022}.  

QV is a benchmarking protocol based on  the execution of a pseudo-random quantum circuit  with a fixed but generic form producing a bitstring $\{x\}$  (Fig.~\ref{QV_PRQC}). QV quantifies the largest random circuit $U$ of equal width $N$ (number of qubits) and depth $d$ (number of layers) that the computer successfully implements:

\begin{equation}
\label{QV}
U = U(d), . . . ,U(2)U(1)
\end{equation}
The ideal output distribution is

\begin{equation}
\label{QV}
p_U (x) = |\bra{x}U\ket{0}|^2
\end{equation}
where $\{x\}$ is an observable bit string. 

Benchmarking the QV, one runs circuits with an increasing number of cycles $d=1, ...., d_{max}$ with $d=N$, and measures the success rate for increasing the depth $d$ until one reaches a prescribed success threshold. 
To define when a model circuit U has been successfully implemented in practice, Cross et al. \cite{Cross2019} use the heavy output generation (HOG) problem formulated by Aaronson and Chen  \cite{Aaronson2017}: "Given as input a random quantum circuit C (drawn from some suitable ensemble), generate output strings $x_1, ....., x_k$, at least a $2/3$ fraction of which have greater than the median probability in C's output distribution." 
This means that the set of output probabilities  $p_U (x)$ are sorted in ascending order of probability, and the heavy (high probability) output generation problem is to produce a set of output strings $\{x\}$ such that more than two-thirds are heavy, i.e.  greater than the median probability.

Aaronson and Chen \cite{Aaronson2017} state that: "HOG is easy to solve on a quantum computer, with overwhelming success probability, by the obvious strategy of just running C over and over and collecting k of its outputs", and demonstrate  \cite{Aaronson2017} that HOG is exponentially hard for a classical computer.  The important thing is that the approach  \cite{Aaronson2017} makes no reference to sampling or relation problems. Thus, one can shift focus from sampling algorithms to algorithms that simply estimate amplitudes. 

Pelofske et al. \cite{Pelofske2022}  recently published a guide to the QV: "Quantum Volume in Practice: What Users Can Expect from NISQ Devices". QV provides a standard benchmark to quantify the capability of NISQ devices. Interestingly, the QV values achieved in the tests \cite{Pelofske2022} typically lag behind officially reported results and also depend significantly on the classical compilation effort. This is important to have in mind when popular articles announce quantum computing breakthroughs in terms of higher QV values. 

%%%%%%%%%%%%%%%%%%%%%%%%%%%%%%%%%%%%
%%%%%%%%%%%%%%%%%%%%%%%%%%%%%%%%%%%%

\subsubsection{Relevance of metrics for usefulness} 

The definition of QV: $d = N$,  stops short of benchmarking what is needed for useful applications.
Useful algorithms often require the quantum circuit depth $d$ to be much larger than the width $N$ (number of qubits): $d >> N$. This is typically the case when describing the ground-state energy of a molecule with reasonable accuracy.  For example, a small molecule like HCN can be described (STO-6G basis) with $N=14$ and $d \approx 3000 \approx 200 N$ \cite{Wendin2022}.  Similarly, HCN (6-31G basis) can be described using Qiskit with $N=69$ and $d= 6 \times 10^6 \sim  87000 N$ \cite{Lolur2021}. 
These huge circuit depths can most likely be reduced with improved compilation methods (see e.g. \cite{Wendin2022}), but nevertheless indicate the nature of the problem to perform useful calculations. 

For comparison, instead of using random circuits and XEB or QV/HOG as targets, one can generate specific quantum states showing genuine multipartite entanglement (GME) with sufficient fidelity. Mooney et al. \cite {Mooney2021} investigated multiple quantum coherences of  Greenberger-Horne-Zeilinger (GHZ)  states on 11 to 27 qubits prepared on the IBM Quantum Montreal (ibmq\_montreal) device (27 qubits), applying quantum readout error mitigation and parity verification error detection to the states.
In this way, a fidelity of $0.546 \pm 0.017 > 0.5$ was recorded for a 27-qubit GHZ state, demonstrating rare instances of GME across the full device.

Although this experiment may feel more interesting and useful than testing with random circuits, it nevertheless demonstrates that there is a very low probability for creating a 27 qubit GHZ state. For it to be useful, the GHZ state must be created with 100\% probability to serve as starting point for useful information processing.

%%%%%%%%%%%%%%%%%%%%%%%%%%%%%%%%%%%%%%%%%%%
%%%%%%%%%%%%%%%%%%%%%%%%%%%%%%%%%%%%%%%%%%%
%%%%%%%%%%%%%%%%%%%%%%%%%%%%%%%%%%%%%%%%%%%
%%%%%%%%%%%%%%%%%%%%%%%%%%%%%%%%%%%%%%%%%%%

\section{Applications}

\subsection{Quantum approximate optimization algorithm - QAOA} 

The Quantum Approximate Optimization Algorithm (QAOA) was proposed as a heuristic variational method for solving NP-hard combinatorial optimization problems on near-term quantum computers \cite{Farhi2014,Farhi2016}, and constitutes one of the most widespread and active current methods for using NISQ computers \cite{Farhi2022,ZhouLukin2020,Morales2020,ZhangB2022,Harrigan2021,Herrman2021,Borle2021,Dupont2022,Chen2022,Sreedhar2022,Willsch2020,Bengtsson2020,Fitzek2021,Lacroix2020,Weggemans2022,Verdon2019,Moussa2022,Patel2022,Bravyi2020,Bravyi2022,Egger2021,Guerreschi 2019,Wurtz2022,Chandarana2022,Hegade2022}.

\begin{figure}
\centering
\includegraphics[width=14cm]{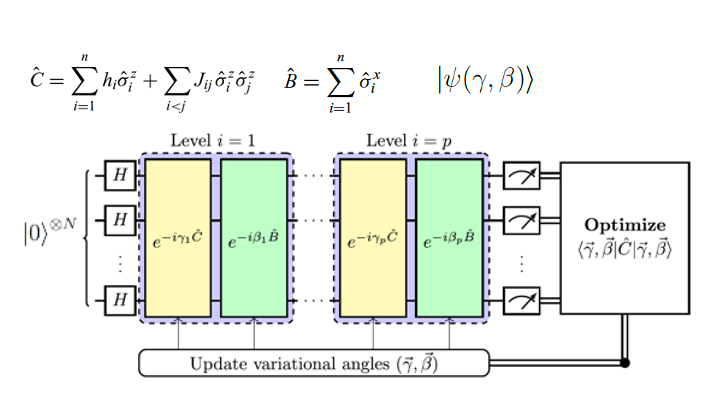} 
\caption{Quantum circuit for the quantum adiabatic optimization algorithm (QAOA). 
The QAOA for a problem specified by the Ising Hamiltonian $\hat C$. 
An alternating sequence of the Ising Hamiltonian $\hat C$ and the transvers mixing Hamiltonian $\hat B$  is applied to an equal superposition of $N$ qubits, producing a trial state function
$ \ket{\psi(\gamma,\beta)}  = \prod_{l=1}^p e^{-i\beta_l \hat B} e^{-i\gamma_l \hat C} \ket{+}^n  $. 
Measurement of the qubit state produces a specific $N$-qubit bitstring, and many repetitions (shots)  of the identical quantum circuit  (loop not shown) creates a distribution used for estimating the cost function  $ \bra{\psi(\gamma,\beta)}C\ket{\psi(\gamma,\beta)}$. A classical optimization algorithm minimizes the cost function by varying the angles
$\gamma,\beta$. The level  $p$ represents the depth of the circuit, determining the number of variational parameters and gates used in the trial function $ \ket{\psi(\gamma,\beta)} $. 
A large circuit width $N$ requires a (very) large depth $p$ for accuracy.
Adapted from \cite {Bengtsson2020}).}. 
\label{QAOA}
\end{figure}

\subsubsection{QAOA basics} 

The QPU prepares a variational quantum state $ \ket{\psi(\gamma,\beta)}$ with $N$ qubits starting from an initial uniform superposition of all possible computational basis states $\ket{+}^{\otimes N}$ generated by Hadamard gates from $\ket{0}^{\otimes N}$ (Fig.~\ref{QAOA}). The second step of QAOA is then to apply an alternating sequence $U = U(p), . . . ,U(2)U(1)\ket{+}^{\otimes N}$ of two parametrized non-commuting quantum gates $U(i) = U(i,\beta_i) U(i,\gamma_i) = e^{-i\beta_l \hat B} e^{-i\gamma_l \hat C}$, followed by measurement generating an $n$-qubit bitstring. Many repetitions (shots) of the same circuit generates a distribution of bitstrings used to evaluate the cost function $ \bra{\psi(\gamma,\beta)}C\ket{\psi(\gamma,\beta)}$. The variational parameters are then updated in a closed loop using a classical optimizer to minimize the cost function.

\subsubsection{QAOA applied to air transportation - tail assignment} 

Industrial optimization has a long history \cite{OngTeo2021}, one of the most famous applications being Toyota's Just-In-Time production system first implemented in 1973 \cite{Monden2012}. 
There is a huge recent literature on optimization for industrial engineering and logistics (see e.g.  \cite{deSousa2019,Csalodi2021,OngTeo2021,Monden2012}), since around 2015 often referred to as Industry 4.0 \cite{Csalodi2021,Rai2021}. 
  
In the following we will discuss one specific example addressing airline scheduling \cite{Wedelin1995}: the performance of the QAOA algorithm for optimizing small but realistic instances of logistic scheduling relevant to airlines.  The problem addressed is called Tail Assignment (TAS) \cite{Gronkvist2005,Svensson2021,Vikstal2020}, assigning individual aircraft (identified by the number on its tail fin) to particular routes,  deciding which individual aircraft  (tail) should operate each flight.

A full approach to TAS is discussed in detail by Svensson et al.  \cite{Svensson2021}, separating TAS into a generation problem and a selection problem.  
In this way, the complex rules only affect the generation problem, whereas the selection problem is often a pure Set Cover or Set Partitioning problem.

The TAS generation problem is responsible for generating the complex aircraft routes.
A flight is a connection between two airports.
A set of flights operated in sequence by the same aircraft (tail) is called a route \cite{Vikstal2020}.
To formulate the TAS problem, let $F$ denote the set of flights  $f$, $T$ the set of tails  $t$, and $R$ the set of all legal routes  $r$. 
In order for a route to be considered legal to operate, it needs to satisfy a number of constraints.

In a full problem description one would include various costs, like the cost of flying a route and the cost of leaving a flight unassigned. In the  decision version of TAS,  the goal is to find any solution satisfying all the constraints, disregarding the costs. 
 Essential aspects of the full TAS selection problem can then be reduced to an Exact Cover decision problem with the constraint
\begin{equation}
\label{TAS}
\qquad \sum_{r \in R} a_{fr} x_r = 1; \;\;\;\; x_r  \in \{0, 1\}
\end{equation}
The constraint matrix \{$a_{fr}$\}   defines the relationship between  $F$ and  $R$ and tells whether a flight $f$ is included in route $r$: $a_{fr}=$ 1 if flight $f$ is covered by route $r$ and 0 otherwise. 
Given the  {\em generated constraint matrix} \{$a_{fr}$\}, the solutions for the decision variable $x_r $  will follow from the solution of the Exact Cover decision problem: $x_r = 1$ if route $r$ should be used in the solution, and is 0 otherwise. 

The constraint can be turned into a cost function
\begin{equation}
\label{Tail}
C = \qquad \sum_{f\in F}(\qquad \sum_{r \in R} a_{fr} x_r - 1)^2
\end{equation}
that can be converted to the classical QUBO model - Quadratic Unrestricted Binary Optimization, which then maps over onto the quantum  Ising model \cite{Glover2022a,Glover2022b,Ajagekar2022} . In the final cost function, the Ising Hamiltonian, the constants (external field and spin interactions) are then determined by the constraint matrix \{$a_{fr}$\} (Eq.~\ref{TAS}). 

Vikstål et al. \cite{Vikstal2020} reduced real-world instances obtained from real flight scheduling to instances with 8, 15 and 25 decision variables, which could be solved using QAOA on a quantum computer laptop simulator with 8, 15 and 25 qubits (routes) respectively. For these small instances, the problem was reduced to an exact cover problem with one solution in each instance. 

The same TAS problem was studied by Willsch et al. \cite{Willsch2022a}  mapped onto a 40-qubit problem and 472 flights. For each of the 40 routes, the constraint matrix defines all flights that are covered by this route. 
As explained, the exact cover problem is to find a selection of routes (i.e., a subset
of rows of the constraint matrix) such that all 472 flights are covered exactly once. The exact cover problem was programmed on D-Wave Advantage and 2000Q. The problem instance has the unique ground state $\ket{0000000001010010011001000001000000000110}$, where each qubit represents a flight route. The ground state contains nine 1's meaning that for this particular instance, the solution consists of nine routes. Each route is assigned to an aircraft. All other states represent invalid solutions, in the sense that not all 472 flights are covered exactly once.

%%%%%%%%%%%%%%%%%%%%%%%%%%
%%%%%%%%%%%%%%%%%%%%%%%%%%
%%%%%%%%%%%%%%%%%%%%%%%%%%
%%%%%%%%%%%%%%%%%%%%%%%%%%

\subsection{Variational quantum eigensolver - VQE} 

\subsubsection{VQE basics }

\begin{figure}
\includegraphics[width=16cm]{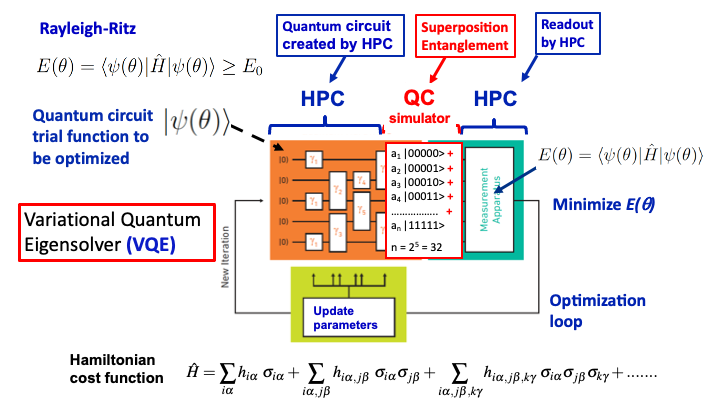} 
\caption{The Variational Quantum Eigensolver (VQE) implements the Rayleigh-Ritz variational principle: $E(\theta) = \bra{\psi(\theta)} \hat H \ket{\psi(\theta)}  \ge E_0$. The trial function $\ket{\psi(\theta)}$ is created as a quantum circuit by the HPC and sent to the backend QPU (or simulator). The gates are then executed in the QPU by pulses generated by the classical electronics and software control system in the stack. The quantum register is finally measured, producing a bitstring used to evalute the Hamiltonian cost function  $\hat H$. This is repeated many times to evaluate the energy  $\bra{\psi(\theta)} \hat H_i \ket{\psi(\theta)} $ with sufficient accuracy. The optimization loop then updates the parameters $\{\theta_i\}$  to minimize the energy. 
}
\label{VQE}
\end{figure}

The Variational Quantum Eigensolver (VQE) implements the Rayleigh-Ritz variational principle (Fig.~\ref{VQE})  \cite{Wendin2017,Lee2019}:
\begin{equation}
\label{RR}
E(\theta) = \bra{\psi(\theta)} \hat H \ket{\psi(\theta)}  \ge E_0 
\end{equation}
The VQE is a classical-quantum hybrid algorithm where the trial function $\ket{\psi(\theta)}$ is created in the qubit register by gate operations. Calculating the expectation value on a QPU, the energy is estimated via quantum state tomography of each of the Pauli operator products of $\hat H$. In quantum simulations on an HPC, the state vector is available classically, and the expectation value of H can be evaluated directly. The VQE scales badly for large molecules (due to repeated measurements/tomography to form the expectation value of the Hamiltonian, $\av{\hat H}$. Nevertheless, the VQE is the common approach for small molecules with present NISQ QPUs \cite {Cao2019,McArdle2020,McClean2021,Fedorov2022}. The phase-estimation algorithm (PEA) scales better, but involves much deeper circuits, puts much higher demands on the coherence time of the q-register, and needs advanced QEC. 
%

%%%%%%%%%%%%%%%%%%%%%%%%%%%%%%%%%
%%%%%%%%%%%%%%%%%%%%%%%%%%%%%%%%%

\subsubsection{VQE applied to chemistry:}

For an overview of applications to chemistry, see reviews \cite {McArdle2020,McClean2021,Fedorov2022} and specific applications  \cite{Lee2019,Kim2022,Sokolov2020,Elfving2021,Sapova2022,Lolur2021,Wendin2022,Lolur2023,Sugisaki2019,Sugisaki2022,ChenBooth2021,YuZhang2022,JAGT2023,Grimsley2019,Tang2020,Tkachenko2021,Lan2022}. 

In VQE calculations for quantum chemistry \cite{Wendin2017,Lee2019} one typically starts from an ansatz of the quantum state $\ket{\psi(\theta)}  =  U(\theta) \ket{\psi_{ref}}$ with variational parameters $\theta$, where $U(\theta)$ is a unitary operator describing the quantum circuit, and $\ket{\psi_{ref}}$ is the initial state.
$U(\theta)$ could be a heuristic "hardware efficient" quantum circuit \cite{Kandala2017} or a more elaborate unitary coupled cluster (UCC) expansion, with a Hartree-Fock \cite{Wendin2017,Lee2019,Sokolov2020} or multi-configuration \cite{Sugisaki2019,Sugisaki2022}  initial reference states  . 
    
The UCC ansatz of the quantum state $\ket{\psi(\theta)}$:
\begin{equation}
\ket{\psi(\theta)} =  \hat U(\theta)  \ket{\psi_{ref}}  =  e^{T(\theta)-T(\theta)^\dagger}  \ket{\psi_{ref}} 
\label{CC}
\end{equation}
 can be expanded: 
\begin{equation}
T(\theta) = T_1 + T_2 + T_3 + .... + T_N    
\label{CCT}
\end{equation}
producing $1,2,3, ...., N$ electron-hole excitations from the N-electron reference state.  
The first two terms
\begin{eqnarray}
T_1 =   \sum_{pq} t(\theta)_{pq} \; c^\dagger_p c_q \label{T1}; \;\;\; 
T_2 = \sum_{pqrs} t(\theta)_{pqrs} \; c^\dagger_p c^\dagger_q c_r c_s \label{T2} 
\label{CCT1T2}
\end{eqnarray}
with fermionic creation ($c^\dagger_i$) and annihilation ($c_i$) operators generate single (S) and double (D) excitations and produce the parametrized UCCSD trial-state approximation. In particular $ t(\theta)_{pq} = \theta_i$ and $t(\theta)_{pqrs} = \theta_j$ for all combinations of the indices $pqrs$. 

The trial-state fermionic operator $U(\theta)$ must now be mapped onto qubit spin operators. Common transformations (codings) are Jordan-Wigner (JW), Bravyi-Kitaev (BK) and Parity, all designed to impose the anticommutation rules. 
In the case of the  UCC ansatz, the exponential is expanded into exponentials of large numbers of products of Paul spin-operators acting on qubits. The initial trial state is then constructed through entangled quantum circuits: combinations of parametrized 1q-rotation gates and entangling 2q gates. The size of the quantum circuit can finally be reduced by qubit reduction schemes. All this results in a state vector for the trial state.

The fermionic operators $c^\dagger_i$ and $c_i$ in the molecular Hamiltonian
\begin{equation}
\hat H =   \sum_{pq} h_{pq}c^\dagger_p c_q + \frac{1}{2}\sum_{pqrs} h_{pqrs}c^\dagger_p c^\dagger_q c_r c_s 
\label{Atom}
\end{equation}
 must also be expanded in products of Pauli spin-operators using codings like JW, BK or Parity, resulting in the generic interaction form:
\begin{equation}
\hat H =   \sum_{i\alpha} h_{i\alpha}\; \sigma_{i\alpha}   + \sum_{i\alpha,j\beta}  h_{i\alpha,j\beta} \;\sigma_{i\alpha}  \sigma_{j\beta} +  \sum_{i\alpha,j\beta,k\gamma}  h_{i\alpha,j\beta,k\gamma} \;\sigma_{i\alpha}  \sigma_{j\beta} \sigma_{k\gamma}  + .......
\label{Atom_Pauli}
\end{equation}
where $\sigma_{i\alpha}$ corresponds to the Pauli matrix $\sigma_{\alpha}$ for $\alpha \in \{0,x,y,z\}$, acting on the $i$-th qubit. 
The expectation value  $\av{\hat H}$ can then be calculated in two ways: (1) State-vector approach: direct calculation of  by matrix operations; (2) Measurement approach: generating an ensemble of identical trial states and measuring the Pauli operators of the Hamiltonian terms $\hat H_i$.  
 
 The original UCC exponential (Eq.~\ref{CC}) is then expanded into exponentials of large numbers of products of Paul spin-operators acting on qubits:
$ e^{-i \theta \sigma_{1z} \sigma_{2z}} $; $ e^{-i \theta \sigma_{1z} \sigma_{2z}  \sigma_{3z}} $, etc.
The parametrized initial entangled quantum circuit  $U(\theta)$ for the UCCSD trial state is then finally constructed through  combinations of parametrized one-qubit rotation gates and entangling two-qubit CNOT gates, resulting in a state vector $\ket{\psi(\theta)}  =  U(\theta) \ket{\psi_{HF}}$ for the trial state. 

 Lolur et al.  \cite{Lolur2021} have benchmarked the VQE as implemented in the Qiskit software package on laptops and HPCs \cite{LolurHPC2021}, applying it to computing the ground state energy of water, H$_2$O, hydrogen cyanide, HCN, and a number of related molecules. The energies have been determined using the  Qiskit statevector backend to directly calculate  $ \bra{\psi(\theta)}\hat H\ket{\psi(\theta)}$ through matrix multiplication rather than repeated measurement.
Clearly, substantial classical computational resources are needed to compute these systems on classical HPC quantum simulators. 
It is evident that for problems with QChem-inspired ans\"atze, even small numbers of qubits lead to large numbers of gates. And this is then amplified by the variational procedure with many parameters and iterations. The large number of gates will severely limit the types of molecules that can be used for benchmarking real quantum HW. And it will also limit what can be simulated on HPC quantum simulators. QChem problems will provide serious challenges and benchmarks for testing HPC and quantum HW NISQ implementations.

To utilize VQE and achieve near chemical accuracy will be extremely challenging for NISQ processors. 
It is problematic or impossible to achieve chemical accuracy with conventional HPC VQE-simulators already for small molecules such as HCN. But, there is no way around it: one must  benchmark and challenge existing quantum HW and SW with available resources. The water molecule is a kind of "gold standard" even for forefront HPC applications, and H$_2$O is an excellent candidate for testing the VQE on quantum HW. 
Nevertheless, at the present stage of the NISQ era, one has to start with "easy" applications and simple approximations just to benchmark the quantum HW.

The concept of "hardware-efficient" trial functions  \cite{Kandala2017} is an attempt to short-circuit systematic UCCSD approaches and still introduce essential electron correlation.
The recently developed adaptive VQE  \cite{Grimsley2019,Tang2020} and related further developments \cite{Tkachenko2021,Lan2022} provide a more systematic approach to including electron correlation processes in order of monotonically decreasing  weight.
Nevertheless, the electron-correlation problem is computationally hard (NP-hard), so there is no easy way around it. State-of-the-art HPC computation of accurate molecular energies based on the Schr\"odinger equation defines the resources needed, and they are indeed huge  \cite{Calvin2021,Jones2022}.  

HPC quantum simulators cannot be more efficient than systematic HPC brute-force Full CI calculations. 
Quantum advantage will be possible by definition as soon as a quantum register exceeds the available RAM memory of an HPC. But to profit from that potential quantum advantage, the QPU must be able to run the q-algorithm to solution, and that will involve a very large number of gate operations even for the VQE. So, this is the ultimate challenge of the NISQ era.

%%%%%%%%%%%%%%%%%%%%%%%%%%%%%%%%%%%%%%%%%%%
%%%%%%%%%%%%%%%%%%%%%%%%%%%%%%%%%%%%%%%%%%%
%%%%%%%%%%%%%%%%%%%%%%%%%%%%%%%%%%%%%%%%%%%

\subsection{Simulating physical systems on engineered quantum platforms} 

Feynman's original idea was to simulate quantum systems with engineered quantum systems. 
Quantum simulation with analog quantum circuits tunes the interactions in a controllable quantum substrate to describe the Hamiltonian of the system to be simulated, and then anneals the systems toward their ground states by lowering the temperature. 

A comprehensive review \cite {Daley2022} describes how quantum simulation can be performed  already today through special-purpose analog quantum simulators, arguing that a first practical quantum advantage already exists in the case of specialized applications of analog devices. A particular example is quantum simulation of 2D antiferromagnets with hundreds of Rydberg atoms trapped in an array by optical tweezers \cite{Scholl2021}, and Lamata et. al. \cite{Lamata2018} and Yu et al. \cite {Yu2022} have developet digital-analog quantum simulation for superconducting circuits. 

The recent development of large-scale superconducting arrays makes it possible to design qubit circuits that simulate a specific physical device, and perform experiments on it \cite{Arute2020a,Neill2021,Gong2021,Mi2022,PZhang2022,Karamlou2022,Landsman2019,Touil2020,Mi2021,Braumuller2021}. In this way Arute et al. \cite{Arute2020a} simulated  separation of the dynamics of charge and spin in the Fermi-Hubbard model,  Neill et al. \cite{Neill2021} simulated the electronic properties of a quantum ring created in the Sycamore substrate, and Mi et al. \cite{Mi2022} investigated discrete time crystals in an open-ended, linear chain of 20 superconducting transmon qubits that were isolated from the two-dimensional Sycamore grid.

We will now discuss a few recent experiments addressing transport, information scrambling and scarring in quantum circuits.

%%%%%%%%%%%%%%%%%%%%%%%%%%%%%%%%%%%%%%%%%%%
%%%%%%%%%%%%%%%%%%%%%%%%%%%%%%%%%%%%%%%%%%%

\subsubsection{Quantum transport and localization:}

Tight-binding lattice Hamiltonians are canonical models for particle transport and localization phenomena in condensed-matter systems.  
To study the propagation of entanglement and observe Anderson and Wannier-Stark localization, Karamlou et al. {\cite {Karamlou2022} experimentally investigate quantum transport in one- and two-dimensional tight-binding lattices, emulated by a fully controllable 3 x 3 array of superconducting qubits in the presence of site-tunable disorder strengths and gradients.

The dynamics are hard to observe in natural solid-state materials, but they can be directly emulated and experimentally studied using engineered quantum systems. The close agreement between the experimental results, simulations, and theoretical predictions  \cite {Karamlou2022} results from high-fidelity, simultaneous qubit control and readout, accurate calibration, and taking into account the relevant decoherence mechanisms in the system.

Karamlou et al. \cite {Karamlou2022} emphasize that although the experiments are performed on a small lattice that can still be simulated on a classical computer, they demonstrate a platform for exploring larger, interacting systems
where numerical simulations become intractable.

\subsubsection{Quantum information scrambling:}

Quantum scrambling is the dispersal of local information into many-body quantum entanglements and correlations distributed throughout an entire system,  leading to the loss of local recoverability of quantum information \cite{Landsman2019,Touil2020,Mi2021,Braumuller2021}. 

Following \cite{Mi2021}, the approach is based on measuring the out-of-time-order correlator (OTOC)   $C = \bra{} \hat O^*(t)\hat M^*\hat O(t)\hat M\ket{}$ between a unitary local perturbation operator $\hat O(t)$ and unitary operator $\hat M$ which is a Pauli operator on a different
qubit.  Scrambling means a local perturbation is rapidly amplified over time. During the time evolution, $\hat O(t)$ becomes increasing nonlocal, which leads to decay of correlation function due to the spreading of the excitation all over the system.

The perturbation operator can be modeled as $\hat O(t)=\sum_i w_i \hat B_i$, where $\hat B_i=b_1(i)   \otimes b_2(i)   \otimes b_3(i)  ....   $ is a string of single-qubit basis operators acting on different qubits, and $w_i$ are the weights of the operator strings.
Scrambling involves two different mechanisms: (i) Operator spreading, and (ii) generation of operator entanglement. 
Operator spreading (i) means that the strings of single-qubit basis operators $\hat B_i$  get expanded, spreading over more qubits, while (ii) generation
of operator entanglement is reflected in the growth in time of the minimum
number of terms needed to expand  $\hat O(t)=\sum_i w_i \hat B_i$  with a broad distribution of coefficients $w_i$. 

By measuring the OTOC, Mi et al.  \cite{Mi2021}  experimentally investigated the dynamics of quantum scrambling on a 53-qubit Sycamore quantum processor. Engineering quantum circuits that distinguished between operator spreading and operator entanglement, they showed that operator spreading is captured by an efficient classical model, while operator entanglement in idealized circuits requires exponentially scaled computational resources to simulate. 
However, the quantum-supremacy discussion of the influence of noise, making possible classical simulation of large noisy QPUs, suggests that the noise level needs to be reduce substantially before exponentially scaled computational resources are needed.

Recently Braum\"uller et al. \cite {Braumuller2021} also  probed quantum information propagation with out-of-time-ordered correlators (OTOC).
They implemented a $3 × 3$ two-dimensional hard-core Bose--Hubbard lattice
with a superconducting circuit, studied its time reversibility, and measured out-of-time-ordered correlators.
The method \cite {Braumuller2021} relies on the application of forward and backward time evolution steps implemented by interleaving blocks of unitary time evolution and single-qubit gates. 
Extracting OTOCs made it possible to study quantum information propagation in lattices with various numbers of particles, enabling observation of a signature of many-body localization in the 2D hard-core Bose--Hubbard model.

Braum\"uller et al. \cite {Braumuller2021}  propose that  applying the technique to larger lattices may improve our understanding of quantum thermodynamics and black-hole dynamics, as well as of using many-body systems for quantum memories. In addition, experimentally accessing OTOCs in large quantum circuits may provide a powerful benchmarking tool to study future quantum processors. But again, here noise will likely become an issue.

\subsubsection{Many-body Hilbert space scarring: }

Zhang et al. \cite{PZhang2022} have studied many-body Hilbert space scarring (QMBS) on a superconducting processor. QMBS is a weak form of ergodicity breaking in strongly interacting quantum systems, meaning that the system does not visit all parts of phase space. This presents opportunities for mitigating thermalization-induced decoherence due to scrambling (eignestate thermalization hypothesis, ETH) in quantum information processing applications. 

Utilizing a programmable superconducting processor with 30 qubits and tunable couplings, Zhang et al. \cite{PZhang2022} create Hilbert space scarring in a non-constrained model in different geometries, including a linear chain and quasi-one-dimensional comb geometry, by approximately decoupling from the qubit substrate. 
By reconstructing the full quantum state through quantum state tomography on four-qubit subsystems, they provide strong evidence for QMBS states by measuring qubit population dynamics, quantum fidelity and entanglement entropy after a quench from initial unentangled states. The QMBS is found to be robust to various imperfections such as random
cross-couplings between qubits, and it persists beyond 1D systems.

The experimental findings also broaden the realm of scarring mechanisms and identify correlations in QMBS states for quantum technology applications.
Comparing with other qubit platforms, Zhang et al. \cite{PZhang2022} state that the superconducting platform can process the same quantum information in a shorter time, implying advantages of QMBS in a superconducting platform for more practical quantum-sensing and metrology applications.

%%%%%%%%%%%%%%%%%%%%%%%%%%%%%%%%%%%%%%%%%%%
%%%%%%%%%%%%%%%%%%%%%%%%%%%%%%%%%%%%%%%%%%%
%%%%%%%%%%%%%%%%%%%%%%%%%%%%%%%%%%%%%%%%%%%
%%%%%%%%%%%%%%%%%%%%%%%%%%%%%%%%%%%%%%%%%%%
%%%%%%%%%%%%%%%%%%%%%%%%%%%%%%%%%%%%%%%%%%%

\section{Key issues}

%%%%%%%%%%%%%%%%%%%%%%%%%%%%%%%%%%%%%%%%%%%
%%%%%%%%%%%%%%%%%%%%%%%%%%%%%%%%%%%%%%%%%%%
%%%%%%%%%%%%%%%%%%%%%%%%%%%%%%%%%%%%%%%%%%%

\subsection{Noise and loss of information - a common experience.}

A common classical experience might illuminate what quantum computing is facing in the present NISQ era.
Onboard an airplane, listening to music using the cheap versions of headphones offered by airlines in economy class can be a less-than-satisfactory  experience.  
To start with, the headphone sound emitters have narrow bandwidth and large distortion, certainly not improving the limited quality or the source itself.
Then the high-frequency background noise in the cabin from the air conditioning and engines may swamp the music signal. And finally, the sensitivity and frequency response of passenger's ears, and the processing in the brain \cite{CPS,CAPD,Souffi2023}, may be less than perfect, making it difficult to discriminate against the noise. For the traveller, high-quality headphones with noise suppression therefore make a big difference. Then the external noise from the environment is processed in real time: recorded, inverted, and subtracted. 

This is a useful analogy in the case of a single qubit in a noisy environment. However, in order to describe the influence of noise on a multi-qubit processor, one might better  illustrate the situation in terms of the "Cocktail Party Syndrome" \cite{CPS}, referring  to the difficulty to entertain a meaningful conversation within a group of people in a very noisy environment. Here, also simultaneous "two-body interactions" between members of the group add "correlated" noise to the "random" noise from the background. 

In our classical example, one can informally define error suppression, error mitigation, and error correction: 
Error suppression means creating high-quality hardware: the signal input is perfect; the classical bits are perfect; the sound generators in the headphones are perfect.  
Error mitigation means eliminating background cocktail party (channel) noise, e..g. via noise inverting devices, as well as eliminating noise generated within the group (e.g. noise within the brain from alcohol consumption; tinnitus; etc.). Error mitigation would also include recording of the session and providing a clean edited transcript afterwords (post-processing). 
Error correction means coding the information such that any errors can be traced and corrected, in real time or at the end.

When it comes to real quantum computers, similar concepts and actions apply: one talks about quantum error suppression (QES), quantum error mitigation (QEM), and quantum error correction (QEC).

%%%%%%%%%%%%%%%%%%%%%%%%%%%%%%%%%%%%%%%%%%%
%%%%%%%%%%%%%%%%%%%%%%%%%%%%%%%%%%%%%%%%%%%
%%%%%%%%%%%%%%%%%%%%%%%%%%%%%%%%%%%%%%%%%%%

\subsection{Fighting imperfections and noise in quantum processors}

Key issues concern the impact of imperfections and noise on computational capacity. NISQ devices are noisy, which creates decoherence and computational errors due to qubit relaxation and dephasing. One talks about three main types of noise: incoherent, coherent, and correlated.
JJ qubits are embedded in a sea of fluctuating defects creating stochastic charge fluctuations -- incoherent noise -- capable of driving unwanted qubit transitions causing relaxation.  Moreover, qubit control via microwave waveguides and magnetic flux lines is subject to stochastic fluctuations influencing the precision of quantum operations. These fluctuating fields may lead to systematic miscalibrations, drift, and crosstalk -- coherent noise --  that  in principle can be reversed \cite{Krantz2019,Hashim2021,Ahsan2022}. Finally, recent work shows that impact of cosmic rays can generate quasiparticles that create correlated charge noise and qubit relaxations on a length scale of  hundreds of micrometer \cite{Vepsalainen2020,Wilen2021}.  

In order to mitigate the effects of noise, one talks about quantum error suppression (QES), quantum error mitigation (QEM), and quantum error correction (QEC).

%%%%%%%%%%%%%%%%%%%%%%%%%%%%%%%%%%%%%%%%%%%
%%%%%%%%%%%%%%%%%%%%%%%%%%%%%%%%%%%%%%%%%%%

\subsubsection{Quantum error suppression:}
QES refers to various efforts to maximize the Quantum Volume by improving the fabrication and operation of the quantum hardware (HW). (This excludes active feedback, treated as QEC). 

On the fabrication side, the main issues concern minimizing decoherence \cite{Krantz2019,Burnett2019,Vepsalainen2020,Wilen2021,Spring2022,Muller2019,Bilmes2020,Caroll2022}  and cross talk \cite{PZhao2022,Spring2022}. 
For a given QPU circuit, the issue is  to maximize gate fidelities and speed via optimal control (e.g. pulse shaping) based on advanced characterization of the device \cite{Wittler2021}.

%%%%%%%%%%%%%%%%%%%%%%%%%%%%%%%%%%%%%%%%%%%
%%%%%%%%%%%%%%%%%%%%%%%%%%%%%%%%%%%%%%%%%%%

\subsubsection{Quantum error mitigation:} 
QEM aims to produce accurate expectation values of observables.
It refers to various software methods to alleviate the effects of noise on computational results during execution of an algorithm on a QPU \cite{Temme2017,YLi2017,Kandala2017,Endo2018,Song2019,vandenBerg2022,Takagi2022,Lolur2023}.  

QEM effectively involves creating a noisy distribution of results, and then extracting the desired quantum information via post-processing.  
Currently, the main principles are: zero noise extrapolation (ZNE)  \cite{Temme2017,YLi2017,Kandala2017,Endo2018}, and  probabilistic error cancellation (PEC)  \cite{Temme2017,YLi2017,Endo2018,Song2019,vandenBerg2022,Takagi2022}. 
 The first scheme (ZNE) does not make any assumption about the noise model other than it being weak and constant in time \cite{Temme2017}. 
The second scheme (PEC) can tolerate stronger noise; however, it requires detailed knowledge of the noise model  \cite{Temme2017}.

{\em Zero noise extrapolation, ZNE}, works by physically increasing the impact of noise, determining a curve describing  how the expectation value of some observable varies with noise.  Variation of the noise strength can be done/simulated by varying the 1q- and 2q-gate times. Given enough points to determine the variation, the curve is then extrapolated back to zero noise, providing a best estimate of the expectation value. This has been implemented successfully in a number of experimental and theoretical investigation \cite{Kandala2019,Giurgica-Tiron2020,Schultz2022,Pascuzzi2022}. 

The zero-noise extrapolation requires sufficient control of the time evolution to implement the rescaled dynamics and hinges on the assumption of a large time-scale separation between the dominant noise and the controlled dynamics  \cite{Temme2017}.

{\em Probabilistic error cancellation (PEC)} works by measuring the noise spectrum and applying an inverted  quasi-probability distribution to the result of the computation via post-processing \cite{Temme2017,YLi2017,Endo2018,Song2019,vandenBerg2022,Takagi2022}.
PEC requires a full characterization of the noisy computational operations. To obtain this to sufficient precision is challenging in practice  \cite{Temme2017}. Nevertheless, Song et al. \cite{Song2019} experimentally demonstrated that PEC based on a combination of gate set tomography (GST) and quasi-probability decomposition can substantially reduce the error in quantum computation on a noisy quantum device. Moreover, Van den Berg et al. \cite{vandenBerg2022} have presented a practical protocol for learning and inverting a sparse
noise model that is able to capture correlated
noise and scales to large quantum devices, demonstrating PEC on a superconducting quantum processor with crosstalk
errors. 

In contrast, Leyton-Ortega et al.  \cite{Leyton-Ortega2023} present a method to improve the convergence of variational algorithms by replacing the hardware implementation of certain Hermitian gates with their inverses, resulting in
noise cancellation and a more resilient quantum circuit. 
This is demonstrated on superconducting quantum processors running the variational
quantum eigensolver (VQE) algorithm to find the H2 ground-state energy. 

Another QEM method has been developed by Lolur et al. \cite{Lolur2023} for quantum chemical computations on NISQ devices  -- Reference-State Error Mitigation (REM). The method relies on determining the exact error in energy due to hardware and
environmental noise for a reference wavefunction that can be feasibly evaluated on a classical computer. REM is shown to drastically improve the computational accuracy at which total energies of molecules
can be computed using current quantum hardware.

%%%%%%%%%%%%%%%%%%%%%%%%%%%%%%%%%%%%%%%%%%%
%%%%%%%%%%%%%%%%%%%%%%%%%%%%%%%%%%%%%%%%%%%

\subsubsection{Quantum error correction:}

QEC refers to methods to code quantum information into logical qubits that can be measured, errors detected and identified, and logical qubits restored  \cite{Fowler2012,Kelly2015,Terhal2015,Wendin2017,Roffe2019,Bravyi2022,Ryan-Anderson2021,Ryan-Anderson2022,Postler2022,Andersen2020,Marques2022,Chen2021,Krinner2022,Acharya2022,Piveteau2021,Ouyang2021,Suzuki2022}. The initial application of the surface code to superconducting devices \cite{Kelly2015} was followed up  by several groups \cite{Andersen2020,Marques2022,Chen2021}, and has recently resulted in  demonstrations of repeated cycles of QEC. Krinner et al. \cite{Krinner2022} implemented a 17 qubit distance-3 surface code, while Acharya et al. \cite{Acharya2022} implemented a 49 qubit distance-5 code. 

Acharya et al. \cite{Acharya2022} draw particular attention to the question whether scaling up the error-correcting code size will reduce logical error rates in a real device. They answer it by reporting on a 72-qubit superconducting device supporting a 49-qubit distance-5 (d = 5) surface code that narrowly outperforms its average subset 17-qubit distance-3 (d=3) surface code, demonstrating a critical step towards scalable quantum error correction.  

Even though John Martinis' group at USCB in 2014  demonstrated that superconducting quantum circuits were at the surface code threshold for fault tolerance  \cite{Barends2014}, this did not mean  that the route to QEC was a straight one - it is all about scaling. 
When the physical error rate is high, the  logical error probability increases with increasing system size, while sufficiently low physical error rates will  lead to the desired exponential suppression of logical errors.

Acharya et al. \cite{Acharya2022} show that their experiment lies in a crossover regime where increasing system size initially suppresses the logical error rate before, due to finite-size effects, later increasing it.  They estimate that component performance must
improve  significantly to achieve practical scaling. In any case, the work demonstrates the first step in the process of suppressing logical errors by scaling a quantum error-correcting code.

%%%%%%%%%%%%%%%%%%%%%%%%%%%%%%%%
%%%%%%%%%%%%%%%%%%%%%%%%%%%%%%%%
%%%%%%%%%%%%%%%%%%%%%%%%%%%%%%%%

\subsection{Scaling up for practical quantum advantage }

The concept of Quantum Advantage (QA) has emerged as a reaction to the more dramatic notion of Quantum Superiority (QS) \cite{Preskill2012}. In principle, QS is what we need - exponential advantage of over classical computers. However, this is only possible with functional QEC. QA emphasizes enhanced performance relative to specific classical algorithms for real-world use cases, typically addressing variational problems like QAOA and VQE. 

Currently there seems to be two opposite uses of practical QA (PQA):  (i) effectively describing QS in huge QEC machines, and (ii) mainly providing some useful speedup relative to classical algorithms. In the present NISQ era there are essentially two ways to look at the scaling up of QPUs, what we will refer to as {\em QPU-centric} and {\em HPC-centric}.

\subsubsection{QPU-centric approach:}
Here QPUs are scaled up in tune with HW progress to push the limits in experiments testing quantum supremacy \cite{Arute2019a,Wu2021,Zhu2022}, QEC  \cite{Andersen2020,Chen2021,Marques2022,Krinner2022,Acharya2022},  and physics\cite{Barends2015,Barends2016,Kandala2017,Kandala2019,Gong2019,Arute2020a,Arute2020b,Neill2021,Gong2021,Gong2022,PZhang2022,Mi2022,Stanisic2022}.   
The main role of the classical computer is to serve a single QPU with pre- and post-processing for running quantum circuits. The maximum number of qubits in the QPU so far is 72 \cite{Acharya2022}. Further scaling up the number of qubits will make it possible to systematically build larger logical qubits and larger code distances. In a blog post in May 2021  \cite{Lucero2021}, Erik Lucero at Google began with a bold statement: "Within the decade, Google aims to build a useful, error-corrected quantum computer." The key issue in 2029 will most likely be: "useful to whom?" Researchers or industry?

\subsubsection{HPC-centric approach:}
Here the HPC supercomputer seamlessly integrates collections of parallel CPUs, GPUs and QPUs. The quantum big picture is to boost classical performance by including QPU subroutines approximately solving specific NP-hard problems that form classical bottlenecks. In this sense, the IBM roadmap and philosophy are {\em HPC-centric}, even though it is described by IBM as quantum-centric supercomputing \cite{Gambetta2022,Bravyi2022}. 
The maximum number of qubits is currently 433, in the Osprey QPU \cite{Osprey2022}. Osprey will be operated as a system of small parallel QPUs  to achieve a computational advantage in the near term by combining multiple QPUs through circuit knitting techniques \cite{Bravyi2022,Piveteau2022}, improving the quality of solutions through error suppression and mitigation, and focusing on heuristic versions of quantum algorithms with asymptotic speedups. 

The IBM Q Experience has created an ecosystem based on Qiskit, providing a versatile programming and testbed environment \cite{Garcia-Perez2020, Bravyi2022}, beginning to look like an industry standard. 
However, super-polynomial speedup does not belong to the NISQ era, and practical quantum advantage is elusive. Realistically, industrial users will not profit from quantum accelerators in the near term, so how can this large quantum effort be justified? 
The answer seems to be that IBM is going for useful quantum advantage via quantum parallel processing provided by QPU accelerators integrated in an efficient runtime HPC environment. Again the question is: useful to whom?  And is useful quantum advantage possible without QEC?

The Quantum Volume (QV) effectively represents the size of a qubit register for which one can entangle all of its qubits in a single shot. Currently, for IBM the best value is $QV=2^9=512$, corresponding to a 9 qubit quantum circuit 9 levels deep. This means that there is no point running algorithms requiring more than that. Instead, one can configure the operating system to run a number of 9-qubit mini-QPUs in parallel to speed up the rate for creating measured distributions, thus reducing the time to solution for computing expectation values of physical variables, like energy. 

Scale, quality, and speed are three key attributes to measure the performance of near-term quantum computers  \cite{Gambetta2022,Bravyi2022}. In the NISQ  era,  the QPU will spend very little time computing compared with the time spent by the CPU on pre- and post-processing before and after every call to the QPU.  Calls that will be very frequent when solving variational problems. Circuit Layer Operations per Second (CLOPS) \cite{Wack2021} is a measure correlated with how many QV circuits (mini-QPUs) a QPU can execute per unit of time, therefore shortening the time to solution. 

At the IBM  Summit 2022, Jay Gambetta \cite{Gambetta2022b} pledged that in 2024 IBM will offer a system that will generate reliable outcomes running 100 qubits with gate depth of 100: "So creating this $100×100$ device will really allow us to set up a path to understand how can
we get quantum advantage in these systems and lay a future going forward."
It must be noted, however, that this does not mean executing a 100q quantum circuit with depth 100 coherently "in a single shot", achieving $QV=2^{100}$ - that would be a quantum-earth shaking demonstration of quantum superiority.

Microsoft has developed a framework for quantum resource estimation \cite{Beverland2022}, to estimate resources required across the stack layers for large-scale quantum applications, finding (as expected) that hundreds of thousands to many millions of
physical qubits are needed to achieve practical quantum advantage.
Beverland et al. \cite{Beverland2022} maintain that the best solution is a monolithic QPU with, say, 10 million controllable, fast, and small qubits. The stack control system must be able to run millions of parallel high-fidelity gates at high speed, as well as reading out millions of qubits in parallel.

Not surprisingly, no qubit technology currently implemented satisfies all of these requirements. However, Microsoft suggests that the
recent proposals of electro-acoustic qubits \cite{Chamberland2022}, and the topological qubit approach based on Majorana Zero Modes \cite{Karzig2017} might do it in future.
Practical quantum advantage is on the horizon but needs to be accelerated through a variety of breakthrough techniques, 
The Microsoft view is that  these research directions can be best studied in the context of resource estimation to unlock quantum at scale.

%%%%%%%%%%%%%%%%%%%%%%%%%%%%%%%%%%%%%%%%%%%
%%%%%%%%%%%%%%%%%%%%%%%%%%%%%%%%%%%%%%%%%%%
%%%%%%%%%%%%%%%%%%%%%%%%%%%%%%%%%%%%%%%%%%%

\subsection{Useful NISQ digital quantum advantage -  mission impossible?}

The short answer is: yes, unfortunately probably mission impossible in the NISQ era. %
%The primary reason is that useful QA needs QEC. 
%Moreover, it is far from clear for which algorithms exponential speedup is available in the end. The notion of quantum chemistry as the killer application \cite{Reiher2017} because of exponential advantage is no longer obvious \cite{Liu2022,Lee2022}.  \\
There are two fundamental questions:
(1)  Does the physical problem itself provide a quantum advantage? And (2), does a quantum algorithm have any advantage over a corresponding classical algorithm?

Both question were  the orignal drivers of QC: the exponential advantage of Shor's algorithm for factorization into prime numbers. However, the need for QEC put that problem far in the future. Instead, Matthias Troyer and coworkers \cite{Reiher2017}  promoted quantum chemsitry for catalysts as the most useful killer application motivating the quest for scaling up QC. 
The paper argues that "quantum computers will be able to tackle important problems in chemistry without requiring exorbitant resources", but at the same time concludes that  "The required space and time resources for simulating FeMoco  are comparable to that of Shor's factoring algorithm.  Berry et al. \cite{Berry2019} improved on those results, obtaining circuits requiring less surface code  resources, despite using a larger and more accurate active space. Nevertheless, also this needs extensive QEC and is far beyond NISQ computers.
Liu et al.  \cite{Liu2022} further elaborate on the potential benefits of quantum computing in the molecular sciences, i.e., in molecular physics, chemistry, biochemistry, and materials science, emphasizing the competition with classical methods that will ultimately decide on the usefulness of quantum computation for molecular science.
Lee et al. \cite {Lee2022} have examined the case for the exponential quantum advantage (EQA) hypothesis for the central task of ground-state determination in quantum chemistry.  Key for EQA is for the quantum state preparation to be exponentially easy compared to classical heuristics, which is far from clear and perhaps not even likely \cite{Wang2021,Bittel2021}. Identifying relevant quantum chemical systems with strong evidence of EQA remains an open question \cite {Lee2022}.

The second question: "does a quantum algorithm have any advantage over a corresponding classical algorithm?" is currently a hot topic.
Understanding whether e.g. quantum machine learning (QML) algorithms present a genuine computational advantage over classical approaches is extremely challenging. It seems that quantum inspired classical algorithms 
"dequantizating" quantum algorithms \cite{EwinTang2021,EwinTang2022,Cotler2021,Gharibian2022}  can compete in polynomial time as long as one is not demanding exponentially accurate results. 

Tang and coworkers \cite{EwinTang2022} developed a dequantization framework for analysing QML algorithms to produce formal evidence against exponential quantum advantage. These are fully classical algorithms that, on classical data, perform only polynomially slower than their quantum counterparts. The existence of a dequantized algorithm means that its quantum counterpart cannot give exponential speedups on classical data, suggesting that the quantum exponential speedups are simply an artifact of state preparation assumptions. QML has the best chance of achieving large speedups whenever classical computation
cannot get access to this data (which occurs when input states come from quantum circuits and other physical quantum systems).
This does not yet rule out the possibility of large polynomial speedups on classical data, which could still lead to significant performance improvements in practice with sufficiently good quantum computers  \cite{EwinTang2022}. 

Lloyd et al. \cite{Lloyd2016}, however, proposed an algorithm for topological data analysis (TDA) that could not to be directly
dequantized using the same techniques, raising the question whether a greater speedup was possible with TDA algorithms. This question has now been analyzed in depth by Berry et al. \cite{Berry2022}, proposing a dequantization of the quantum TDA
algorithm which shows that having exponentially large dimension and Betti number is necessary for super-polynomial advantage. The speedup is quartic, which will not be killed by QEC overhead \cite{Babbush2021}.

Based on that, Berry et al. \cite{Berry2022} estimate that tens of billions of Toffoli gates will be suffcient to estimate a Betti number
that should be classically intractable. This number of Toffoli gates is considered to be reasonable for early generations of fully
fault-tolerant quantum computers \cite{Berry2022}, falling somewhere in between quantum chemistry applications and Shor's algorithm in terms of the resources required for quantum advantage.

How this goes together with the very recent result presented by Akhalwaya et al. \cite {Akhalwaya2022} remains to be understood. Quoting the authors: "NISQ-TDA, the first fully implemented end-to-end quantum machine learning algorithm needing only a linear circuit-depth, that is applicable to non-handcrafted high-dimensional {\em classical data}, with potential speedup under stringent conditions. The algorithm neither suffers from the data-loading problem nor does it need to store the input data on the quantum computer explicitly. Our approach
includes three key innovations: (a) an efficient realization of the full boundary operator as a sum of Pauli operators;
(b) a quantum rejection sampling and projection approach to restrict a uniform superposition to the simplices of the
desired order in the complex; and (c) a stochastic rank estimation method to estimate the topological features in the
form of approximate Betti numbers. We present theoretical results that establish additive error guarantees for NISQTDA,
and the circuit and computational time and depth complexities for exponentially scaled output estimates, up
to the error tolerance. The algorithm was successfully executed on quantum computing devices, as well as on noisy
quantum simulators, applied to small datasets. Preliminary empirical results suggest that the algorithm is robust to
noise."

Reconnecting here to quantum chemistry, Gharibian and Le Gall \cite {Gharibian2022} 
have shown how to design classical algorithms that estimate, with constant precision, the singular values of a sparse matrix, implying that the ground state energy in quantum chemistry can be solved efficiently with constant precision on a classical computer. 
However, Gharibian and Le Gall also prove that with inverse-polynomial precision, the same problem becomes BQP-complete, suggesting that the superiority of quantum algorithms for chemistry stems from the improved precision achievable in the quantum setting.

Finally, Huang et al. \cite{Huang2022} investigate quantum advantage in learning from experiments that processes quantum data with a quantum computer. That could have substantial advantages over conventional experiments in which quantum states are measured and outcomes are processed with a classical computer. 

Huang et al. prove that quantum machines can learn from exponentially fewer experiments than the number required by conventional experiments. They do that by assuming having access to data obtained from quantum enhanced experiments like quantum sensing systems and stored in quantum memory (QRAM), allowing the QPU to process quantum input data.
 Exponential advantage is shown for predicting properties of physical systems, performing quantum principal component analysis, and learning about physical dynamics.
Huang et al. \cite{Huang2022}: "Although for now we lack suitably advanced sensors and transducers, we have conducted proof-of-concept experiments in which quantum data were directly planted in our quantum processor." 

In the absence of perfect physical qubits, or QEC, quantum memory is a great challenge. Quantum memory may be far away for quantum computing as needed by e.g. Huang et al. \cite{Huang2022}, but it is essential for the development of quantum repeaters for quantum communication networks. Sullivan et al. \cite{Sullivan2022} investigate random-access quantum memory using chirped-pulse phase encoding. The protocol is implemented using donor spins in silicon coupled to a superconducting cavity, offering the potential for microwave random access quantum memories with lifetimes exceeding seconds.

%%%%%%%%%%%%%%%%%%%%%%%%%%%%%%%%%%%%%
%%%%%%%%%%%%%%%%%%%%%%%%%%%%%%%%%%%%
%%%%%%%%%%%%%%%%%%%%%%%%%%%%%%%%%%%%
%%%%%%%%%%%%%%%%%%%%%%%%%%%%%%%%%%%%

\section{Future directions }

%%%%%%%%%%%%%%%%%%%%%%%%%%%%%%%%%%%%%
%%%%%%%%%%%%%%%%%%%%%%%%%%%%%%%%%%%%
%%%%%%%%%%%%%%%%%%%%%%%%%%%%%%%%%%%%

\subsection{Improved and alternative superconducting qubits} 

A recent comprehensive review by Calzona et al. \cite{Calzona2023} describes and analyzes  the basic concepts and ideas behind the implementation of novel superconducting circuits with intrinsic protection against decoherence at the hardware level. The review explains the basics and performance of state-of-the-art transmons and other single-mode superconducting quantum circuits, and goes on to describe multi-mode superconducting qubits, toward the realization of fully protected qubits engineered in systems with more
than one degree of freedom and/or characterized by the presence of specific symmetries. 

Regarding state-of-the-art tantalum-based transmons \cite{Place2021,CWang2022}, Tennant et al. \cite{Tennant2022} performed low-frequency charge-noise spectroscopy on Ta-based transmons and found distinctly different behaviour compared with Al- and Nb-based transmons.
They conclude that the temperature-dependent behavior of the neighboring charge-configuration transitions is caused by jumps between local charge configurations in the immediate vicinity of the transmon. This is in contrast to  Al- and Nb-based transmons which are dominated by a distribution of TLSs giving rise to 1/f noise, apparently ruling out a collection
of TLSs as the basis of the quasi-stable charge offsets in Ta-based transmons.

Very different types of superconducting devices are semiconductor-superconductor hybrid structures containing Andreev bound states (ABS) \cite{WendinShumeiko2021} and topological Majorana zero modes (MZM) \cite{DasSarma2015,Pikulin2021,Aghaee2022}. 
Recently Pikulin et al. \cite{Pikulin2021} developed an experimental protocol (Toplogical Gap Protocol, TGP) to determine the presence and extent of a topological phase with Majorana zero modes in a hybrid semiconductor-superconductor three-terminal device with two normal leads and one superconducting lead. Now  Aghaee et al. \cite{Aghaee2022}  have presented measurements and simulations of InAs-Al hybrid three-terminal devices that are consistent with the observation of topological superconductivity and Majorana zero modes, passing the TGP. Passing the protocol indicates a high probability of detection of a topological phase hosting Majorana zero modes as determined by large-scale disorder simulations and is a prerequisite
for experiments involving fusion and braiding of Majorana zero modes.

%%%%%%%%%%%%%%%%%%%%%%%%%%%%%%%%%%%%%
%%%%%%%%%%%%%%%%%%%%%%%%%%%%%%%%%%%%
%%%%%%%%%%%%%%%%%%%%%%%%%%%%%%%%%%%%

\subsection{Hybrid distributed computing} 

In Sect. 4.3.2 we talked about distributed quantum processing, running many QPU modules in parallel. It is a matter of debate whether future large scale algorithms can be run on monolithic or modular QPUs fitting inside a single fridge, or whether the algorithms have to be distributed over several fridges and locations. Eventually one will be able to work with clusters of quantum computers connected via  local or global quantum networks, but for now this represents a great challenge waiting for great breakthroughs and a quantum infrastructure.

On a smaller scale, Andrew Cleland and collaborators have done some pioneering work to connect superconducting qubit circuits in different fridges connected by a microwave cable \cite{Zhong2021,Yan2022}, producing multi-qubit entanglement, purification and protection in a quantum network. The microwave connection is photonic, but is limited to local clusters. The needed interfaces for long-distance optical connections are emerging  \cite{Chu2020}, but will probably remain research endeavors for quite some time \cite{Arnold2020,Wang2022,Kumar2022}.

On a larger scale, Ang et al. \cite{Ang2022} have developed architectures for superconducting modular, distributed, or multinode quantum computers (MNQC), employing a `co-design' inspired approach to quantify overall MNQC performance in terms of hardware models of internode links, entanglement distillation, and local architecture.
In the particular case of superconducting MNQCs with microwave-to-optical interconnects,
 Ang et al. \cite{Ang2022}, describe how compilers and software should optimize the balance between local gates
and internode gates, discuss when noisy quantum internode links have an advantage over purely
classical links, and introduce a research roadmap for the realization of early MNQCs. 
This roadmap illustrates potential improvements to the hardware and software of MNQCs and outlines
criteria for evaluating the improvement landscape, from progress in entanglement generation to the
use of quantum memory in entanglement distillation and dedicated algorithms such as distributed
quantum phase estimation. 

As a concrete example, DiAdamo et al. \cite{DiAdamo2021} consider an approach for distributing the variational quantum eigensolver (VQE) algorithm over distributed quantum computers with arbitrary number of qubits in a systematic approach to generate distributed quantum circuits for quantum computing. This includes a  proposal for software-based system for controlling quantum systems at the various classical and quantum hardware levels. 
DiAdamo et al. \cite{DiAdamo2021} emphasize that much effort has gone into distributed computing in the classical computing domain. And since the overlap between the fields is high, one can use this knowledge to design robust and secure distributed quantum computers, and as quantum technologies improve, this may become a reality.

%%%%%%%%%%%%%%%%%%%%%%%%%%%%%%%%%%%%%
%%%%%%%%%%%%%%%%%%%%%%%%%%%%%%%%%%%%
%%%%%%%%%%%%%%%%%%%%%%%%%%%%%%%%%%%%

\subsection{Continuous variables - computing with resonators} 

In this field, the resonator modes are the logical qubits, and the transmon qubits provide ancillas for loading and readout. 
The Yale group is leading the development, and has recently demonstrated some decisive breakthroughs \cite{Sivak2022b}.
The name of the game is to construct logical qubits from linear combinations of (already long-lived) resonator states representing the Gottesman-Kitaev-Preskill (GKP) bosonic code, encoding a logical qubit into grid states of an oscillator.
Sivak et al. \cite{Sivak2022b}  demonstrate a fully stabilized and error-corrected logical qubit
whose quantum coherence is significantly longer than that of all the imperfect quantum components
involved in the QEC process, beating the best of them with a coherence gain of $ G \approx 2.3$.
This was achieved by combining innovations in several domains including the fabrication
of superconducting quantum circuits and model-free reinforcement learning \cite{Sivak2022a}. 

To correct for single-photon loss, Kudra et al. \cite{Kudra2022} have implemented two photon
transitions that excite the cavity and the qubit at the same time. The additional degree of
freedom of the qubit makes it possible to implement a coherent, unidirectional mapping between
spaces of opposite photon parity. The successful experimental implementation,
when supplemented with qubit reset, is suitable for autonomous quantum error correction in bosonic
systems, opening up the possibility to realize a range of non-unitary transformations
on a bosonic mode.

For full scale QEC, various groups have recently investigated the concatenation of CV and DV codes, such as concatenating the single-mode GKP code with the surface code.
Instead, Guillaud and Mirrahimi \cite{Guillaud2019} present a 1D repetition code based on a cat code
as the base qubit for which the noise structure is modified in such a way that quantum error correction becomes of similar complexity as classical error correction and can be performed using a simple repetition code. According to \cite{Guillaud2019}, the specific noise structure can be preserved for a set of fundamental operations which at the level of the repetition code lead to a universal set of protected logical gates. 

Regarding scaling up CV resonator technology, Axline et al.  \cite{Axline2018} have experimentally realized on-demand, high-fidelity state transfer and entanglement between two isolated superconducting cavity quantum memories. 
By transferring states in a multiphoton encoding, Axline et al.  \cite{Axline2018} show that the use of cavity memories and state-independent transfer creates the striking opportunity to deterministically mitigate transmission loss with quantum error correction. The results establish a basis for deterministic quantum communication across networks, and will enable modular scaling of CV superconducting quantum circuits.

The size of superconducting 3D microwave resonators makes it challenging to scale up large 3D multi-qubit CV systems. An alternative may be provided by nanomechanical phononic nanostructures  \cite{Chu2020}. Chu et al.  \cite{Chu2017} experimentally demonstrated strong coupling between a superconducting transmon qubit and the long-lived longitudinal phonon modes of a high-overtone bulk acoustic wave disk resonator (HBAR) formed in thin-film aluminium nitride (AlN). 
Recently, von L\"upke et al. \cite{vonLupke2022} demonstrated HBAR parity measurement in the strong dispersive regime of circuit quantum acoustodynamics, providing basic building blocks for constructing acoustic quantum memories and processors. Moreover, Schrinski et al. \cite{Schrinski2022} measured long-lived HBAR Wigner states, monitoring the gradual decay of negativities over tens of microseconds.  Wollack et al.  \cite{Wollack2022} use a superconducting transmon qubit to control and read out the quantum state of a pair of nanomechanical resonators made from thin-film lithium niobate (LN). The device is capable of fast swap operations,  used to deterministically manipulate the nonclassical and entangled mechanical quantum states. This creates potential for feedback-based operation of quantum acoustic processors. 
Finally, Chamberland et al. \cite{Chamberland2022}
have presented a comprehensive architectural analysis for a proposed fault-tolerant quantum computer based on cat codes concatenated with outer quantum error-correcting codes applied to a system of acoustic resonators coupled to superconducting circuits with a two-dimensional layout.

%%%%%%%%%%%%%%%%%%%%%%%%%%%%%%%%%%%%%%%%%%%
%%%%%%%%%%%%%%%%%%%%%%%%%%%%%%%%%%%%%%%%%%%
%%%%%%%%%%%%%%%%%%%%%%%%%%%%%%%%%%%%%%%%%%%

\subsection{Biochemistry and life science - drivers of quantum computing?} 

In computational science there is the well-established method of multiscale modeling \cite {Fish2021} that gave the Nobel Prize in Chemistry in 2014 to Arieh Warshel for modeling biological functions, from enzymes to molecular machines  \cite {Warshel2014}. Multiscale modelling describes methods that simulate continuum-scale behaviour using information derived from computational models of finer scales in the system, down to molecular quantum levels, bridging across multiple length and time scales. It is then natural to consider using a quantum computer to address the case of a quantum system embedded in a multiscale environment.
Cheng et al. \cite {Cheng2020} review these methods and propose the embedding approach as a method for describing
complex biochemical systems, with the parts  treated at different levels of theory and computed with hybrid classical and quantum algorithms. 

Having come this far, we understand however  that many commonly held views on the power of digital quantum computing, especially in NISQ times, are problematic. Cheng et al. \cite {Cheng2020} illustrate this problem : 
"Chemistry is considered as one of the more promising applications to science of near term quantum computing. Recent work in transitioning classical algorithms to a quantum computer has led to great strides in improving quantum algorithms and illustrating their quantum advantage." 

The bottom line is that if one wants to treat biochemical molecules that contain active regions that cannot be properly explained with traditional algorithms on classical computers, then one should not expect any quantum advantage from NISQ quantum computers. That said,  Cheng et al. \cite {Cheng2020} provide a useful overview of how multiscale modeling involving quantum computers is going to enable biomolecular problems to be tackled in the future. 

To this must added healthcare, life science and artificial intelligence in a broad sense. From the huge data bases of cell biology and human diseases one can design network models describing the networks of Life. In particular Barabasi, Loscalzo and collaborators  
\cite {Barabasi2011,LeeLoscalzo2019} developed the science of network medicine, and machine learning is essential for creating  models for therapies that can design and control the action of drugs  \cite {Santos2021,Infante2021}. 

Biochemistry and life science are already, as always, at the focus of high-performance computing, driving the development of exascale and post-exascale supercomputers, experiencing the limitations and bottlenecks. These are topics and areas that would profit immensely from quantum advantage.
Maniscalco et al.  \cite {Maniscalco2022} recently published a forward-looking white paper: "Quantum network medicine: rethinking medicine with network science and quantum algorithms",  and posit that quantum computing may be a key ingredient in enabling the full potential of network medicine,  laying the foundations of a new era of disease mechanism, prevention, and treatment. 

%%%%%%%%%%%%%%%%%%%%%%%%%%%%%%%%%%%%%%%%%%%
%%%%%%%%%%%%%%%%%%%%%%%%%%%%%%%%%%%%%%%%%%%
%%%%%%%%%%%%%%%%%%%%%%%%%%%%%%%%%%%%%%%%%%%

\subsection{Final perspective}

This great vision reflects the mission of the entire field of quantum computing - to achieve the elusive Quantum Advantage.
Fortunately there are very few problems of importance to mankind that rely on the imminent arrival of quantum computers with QA. 
Quantum computers will evolve in ways that seemed impossible not long ago. And they will provide platforms for fantastic experiments explaining deep fundamental physics and quantum information. And to make that available to the science community for experimentation should be the mission of the QC community. 

However, the usefulness beyond classical computing and algorithms is a very different matter. It depends on practical QA, and remains to be established in practical applications. Practical QA always has to be measured against the performance of classical algorithms and computers. 
HPC-people regard QPUs as a kind of GPUs, closely integrated and expected to accelerate the CPUs when handling NP-hard problems. The HPC+QC integration is happening right now, and during the next five years it will be taken to high levels - IBM is one example. For sure this will challenge and boost the development of competitive classical algorithms and dedicated hardware - but useful QA will remain problematic in the NISQ era.

This is amply illustrated by Goings et al. \cite {Goings2022} discussing how to "explore the quantum computation and classical computation resources required to assess the electronic structure of cytochrome P450 enzymes (CYPs) and thus define a classical-quantum advantage boundary". One conclusion  \cite {Goings2022} is that a large classically intractable CYP model simulation may need 5 million qubits and nearly 10 billion Toffoli gates, and may take 100 QPU hours. Another, less surprising, conclusion is that deep classical chemical insight is essential for guiding quantum algorithms and defining the computational frontier for chemistry.

In that light, the most important near-term use of superconducting quantum processors may be to follow Feynman's original idea and create experiments in large superconducting controllable multi-qubit networks that are impossible for classical computers to simulate. 

The next assessment of the future of QC is planned for 2028 \cite{Wendin2028} and then we can perhaps compare notes.

%%%%%%%%%%%%%%%%%%%%%%%%%%%%%%%%%%%%%
%%%%%%%%%%%%%%%%%%%%%%%%%%%%%%%%%%%%

\section *{Acknowledgement}

This work was supported from the Knut and Alice Wallenberg Foundation through the
Wallenberg Center for Quantum Technology (WACQT).

%%%%%%%%%%%%%%%%%%%%%%%%%%%%%%%%%%%%
%%%%%%%%%%%%%%%%%%%%%%%%%%%%%%%%%%%%%
%%%%%%%%%%%%%%%%%%%%%%%%%%%%%%%%%%%%
%%%%%%%%%%%%%%%%%%%%%%%%%%%%%%%%%%%%
%%%%%%%%%%%%%%%%%%%%%%%%%%%%%%%%

\newpage

\section *{References}


\begin{thebibliography}{99}

\bibitem {Wendin2017} 
G. Wendin, Quantum information processing with superconducting circuits: a review, Rep. Prog. Phys. {\bf80} 106001 (2017).

\bibitem {Gu2017} 
Xiu Gu, et al.,
Microwave photonics with superconducting quantum circuits,
Physics Reports  {\bf718-719}, 1 (2017).

\bibitem {Krantz2019}
P. Krantz, et al.,
A quantum engineer's guide to superconducting qubits, 
Appl. Phys. Rev. {\bf6}, 021318 (2019).

\bibitem {Kjaergaard2020}
M. Kjaergaard, et al.,
Superconducting Qubits: Current State of Play,
Annu. Rev. Condens. Matter Phys. {\bf11}, 369 (2020).

\bibitem {Blais2020}
A. Blais, S. M. Girvin, and W. D. Oliver,
Quantum information processing and quantum optics with circuit quantum electrodynamics,
Nature Phys. {\bf16}, 247 (2020).

\bibitem {Blais2021} 
A. Blais, A. L. Grimsmo, S. M. Girvin, and A. Wallraff,
Circuit quantum electrodynamics, 
Rev. Mod. Phys. {\bf93}, 025005 (2021).

\bibitem {Bruzewicz2019} 
C. D. Bruzewicz, et al.,
Trapped-ion quantum computing: Progress and challenges,
Appl. Phys. Rev. {\bf6}, 021314 (2019).

\bibitem {Brown2021} 
K. R. Brown, J. Chiaverini, J. M. Sage, and H. Häffner,
Materials challenges for trapped-ion quantum computers,
Nature Reviews: Materials {\bf6}, 893 /2021).

\bibitem {Daley2022} 
Andrew J. Daley, et al., 
Practical quantum advantage in quantum simulation,
Nature  {\bf607}, 667 (2022).

\bibitem {QuantumManifesto2016} 
Quantum Manifesto (2016); 
https://qt.eu/app/uploads/2018/04/93056\_Quantum-Manifesto\_WEB.pdf

\bibitem {QFlag}
EU Quantum Flagship,
https://qt.eu

\bibitem {Arute2019a} 
F. Arute, et al., Quantum supremacy using a programmable superconducting processor, 
Nature {\bf574}, 505 (2019).

\bibitem {Wu2021}
Y. Wu, et al., Strong Quantum Computational Advantage Using a Superconducting Quantum Processor, 
Phys. Rev. Lett. {\bf127}, 180501 (2021).

\bibitem {Zhu2022}
Q. Zhu, et al.,
Quantum computational advantage via 60-qubit 24-cycle random circuit sampling,
Science Bullletin {\bf67},  240-245 (2022).

\bibitem {Corcoles2020} 
A. D. Corcoles, et al.,
Challenges and Opportunities of Near-Term Quantum Computing Systems, 
Proc. IEEE {\bf108}, 1338 (2020).

\bibitem {Gambetta2022}
J. Gambetta, Expanding the IBM Quantum roadmap to anticipate the future of quantum-centric supercomputing (10 May 2022;
https://research.ibm.com/blog/ibm-quantum-roadmap-2025

\bibitem {Preskill2012} 
J. Preskill, 
Quantum computing and the entanglement frontier, 
arXiv:1203.5813.

\bibitem {Preskill2018} 
J. Preskill, 
Quantum Computing in the NISQ era and beyond, 
Quantum {\bf2}, 79 (2018).

\bibitem {Bravyi2022}
S. Bravyi, et al.,
The future of quantum computing with superconducting qubits,
J. Appl. Phys. {\bf132}, 160902 (2022). 

\bibitem {Lucero2021}
E. Lucero, Google: Unveiling our new Quantum AI campus, May 18, 2021. 
https://blog.google/technology/ai/unveiling-our-new-quantum-ai-campus/

\bibitem {Sanders2020}
Y. R. Sanders, et al., 
Compilation of fault-tolerant quantum heuristics for combinatorial optimization,
PRX Quantum {\bf1}, 020312 (2020).

\bibitem {Babbush2021}
R. Babbush, et al.,
Focus beyond Quadratic Speedups for Error-Corrected Quantum Advantage,
PRX Quantum {\bf2}, 010103 (2021).

\bibitem {Beverland2022}
M. E. Beverland, et al.,
Assessing requirements to scale to practical quantum advantage
arXiv:2211.07629v1.

\bibitem {Barends2014} 
R. Barends et al., Superconducting quantum circuits at the surface code threshold for fault tolerance,
Nature {\bf508}, 500 (2014).

\bibitem {Barends2015} 
R. Barends et al.,  Digital quantum simulation of fermionic models with a superconducting circuit 
Nat. Commun. {\bf6}: 7654 (2015).

\bibitem {Barends2016} 
R. Barends et al., Digitized adiabatic quantum computing
with a superconducting circuit Nature {\bf 534}, 222 (2016).

\bibitem {Kandala2017} 
A. Kandala, et al., Hardware-efficient quantum optimizer for small molecules and quantum magnets, 
Nature  {\bf549}, 242 (2017).

\bibitem {Song2017} 
C. Song, et al.,
10-qubit entanglement and parallel logic operations with a superconducting circuit,
Phys. Rev. Lett. {\bf119}, 180511 (2017).

\bibitem {Kandala2019}
A. Kandala, et al., 
Error mitigation extends the computational reach of a noisy quantum processor, 
Nature {\bf567}, 491 (2019).

\bibitem {Gong2019} 
M. Gong, et al.,
Genuine 12-Qubit Entanglement on a Superconducting Quantum Processor,
Phys. Rev. Lett. {\bf122}, 110501 (2019).

\bibitem {Gong2021}
M. Gong, et al., Quantum walks on a programmable two-dimensional 62-qubit superconducting processor,
Science {\bf372}, 948-952 (2021).

\bibitem {Gong2022}
M. Gong, et al.,
Quantum Neuronal Sensing of Quantum Many-Body States on a 61-Qubit
Programmable Superconducting Processor,
arXiv:2201.05957v1.

\bibitem {Arute2020a} 
F. Arute, et al,
Observation of separated dynamics of charge and spin in the Fermi-Hubbard model;
arXiv:2010.07965v1.

\bibitem {Arute2020b} 
F. Arute, et al., Hartree-Fock on a superconducting qubit quantum computer, 
Science {\bf369}, 1084 (2020).

\bibitem {Neill2021} 
C. Neill, T. McCourt, V. Smelyanskiy,
Accurately computing the electronic properties of a quantum ring,
Nature {\bf594},  508 (2021).

\bibitem {PZhang2022}
P. Zhang et al., Many-body Hilbert space scarring on a superconducting processor, Nature Physics (13 Oct. 2022); https://doi.org/10.1038/s41567-022-01784-9

\bibitem {Mi2022}
X. Mi, et al., Time-crystalline eigenstate order on a quantum processor, Nature {\bf601}, 531 (2022).

\bibitem {Stanisic2022}
S. Stanisic, et al., Observing ground-state properties of the Fermi-Hubbard model using a scalable algorithm on a quantum computer, Nature Communications {\bf13}:5743 (2022).

\bibitem {Andersen2020}
C. K. Andersen, et al.,
Repeated quantum error detection in a surface code,
Nat. Phys. {\bf16}, 875 (2020).

\bibitem {Chen2021}
Z. Chen, et al. 
Exponential suppression of bit or phase errors with cyclic error correction,
Nature {\bf595}, 383 (2021).

\bibitem {Marques2022}
J. F. Marques, et al., Logical-qubit operations in an error-detecting surface code,
Nat. Phys. {\bf18}, 80 (2022).

\bibitem {Krinner2022}
S. Krinner, et al.,
Realizing repeated quantum error correction in a distance-three surface code,
Nature {\bf605}, 669 (2022).

\bibitem {Acharya2022}
R. Acharya, et al.,
Suppressing quantum errors by scaling a surface code logical qubit,
arXiv:2207.06431v2.

\bibitem {QUTAC2021}
QUTAC, Industry quantum computing applications, 
EPJ Quantum Technology{\bf8}:25 (2021).

\bibitem {Kim2022}
I. H. Kim, et al.,
Fault-tolerant resource estimate for quantum chemical simulations: Case study on Li-ion battery electrolyte molecules
Phys. Rev. Research {\bf4}, 023019 (2022). 

\bibitem {Alexeev2021} 
Y. Alexeev et al.,
Quantum Computer Systems for Scientific Discovery, 
PRX Quantum {\bf2}, 017001 (2021)

\bibitem {Altman2021} 
E. Altman, et al.,
Quantum Simulators: Architectures and Opportunities, 
PRX Quantum {\bf2}, 017003 (2021)

\bibitem {Ang2022}
J. Ang, et al.,
Architectures for Multinode Superconducting Quantum Computers,
arXiv:2212.06167v1.

\bibitem {Ezratty2022}
O. Ezratty,
Mitigating the quantum hype (2022); arXiv: 2202.01925.

\bibitem {DasSarma2022}
S. DasSarma, Quantum computing has a hype problem, MIT Technology Review (March 28, 2022),
https://www.technologyreview.com/2022/03/28/1048355/quantum-computing-has-a-hype-problem/

\bibitem {circuitdepth} 
Circutit depth is defined as the longest path between the data input and the output in terms of number of gates. See Sects. 2.4 and 3.1 for examples.


%%%%%%%%%%%%%%%%%%%%%%%%%%%



\bibitem {Wang2018}
D. Wang, O. Higgott and S. Brierley, A Generalised Variational Quantum Eigensolver,
Phys. Rev. Lett. {\bf122}, 140504 (2019)

\bibitem {Cerezo2021}
M. Cerezo, et al., 
Variational quantum algorithms, 
Nature Reviews Physics {\bf3}, 625 (2021).

\bibitem {Tilly2022} 
J. Tilly, et al.,
The Variational Quantum Eigensolver: A review of methods and best practices, Physics Reports {\bf986}, 1-128 (2022)

\bibitem {Bharti2022}
K. Bharti, et al.,
Noisy intermediate-scale quantum algorithms,
Rev. Mod. Phys. {\bf94}: 015004 (2022).

\bibitem {Franca2021}
 D. S. Franca, and R. Garcia-Patron, Limitations of optimization algorithms on noisy quantum devices, Nature Physics {\bf17}, 1221 (2021).

\bibitem {Abhijith2022}
J. Abhijith et al., Quantum Algorithm Implementations for Beginners, ACM
Transactions on Quantum Computing {\bf3}:18 (2022).

\bibitem {Kowalsky2022}
M. Kowalsky, T.  Albash, I. Hen, and D. A. Lidar, 
3-regular three-XORSAT planted solutions benchmark of classical and quantum heuristic optimizers, 
Quantum Sci. Technol. {\bf7}, 025008 (2022). 



%%%%%%%%%%%%%%%%%%%%%%%%%%%%



\bibitem {Boixo2018}
S. Boixo, et al., 
Characterizing quantum supremacy in near-term devices,
Nature Physics {\bf14,} 595 (2018).

\bibitem {Hangleiter2022}
D. Hangleiter, and J. Eisert,
Computational advantage of quantum random sampling,
arXiv:2206.04079v3.

\bibitem {Aaronson2017}
S. Aaronson and L. Chen, Complexity-Theoretic Foundations of Quantum Supremacy Experiments".
In: 32nd Computational Complexity Conference CCC'17. 2017, pp. 1-67; arXiv: 1612.05903 

 \bibitem {Perdnault2019}
 E. Pednault, J. Gunnels, D. Maslov, and J. Gambetta, 
 On "Quantum Supremacy", IBM\_ResBlog\_Oct21\_2019,
 https://www.ibm.com/blogs/research/2019/10/on-quantum-supremacy/ 

\bibitem {Bouland2019}
 A. Bouland, B. Fefferman, C. Nirkhe and U. Vazirani,
On the complexity and verification of quantum random circuit sampling,
Nature Physics {\bf15}, 159 (2019).

\bibitem {Aharonov2022}
D. Aharonov, et al.,
A polynomial-time classical algorithm for noisy random circuit sampling,
arXiv:2211.03999v1.

\bibitem {Brubaker2023}
B. Brubaker,
New Algorithm Closes Quantum Supremacy Window,
Quanta Magazine, January 9, 2023; 
https://www.quantamagazine.org/new-algorithm-closes-quantum-supremacy-window-20230109/.

\bibitem {FengPan2022a}
F. Pan and P. Zhang
Simulation of Quantum Circuits Using the Big-Batch Tensor Network Method,
Phys. Rev. Lett. {\bf128}, 030501 (2022).

\bibitem {FengPan2022b}
F. Pan, K. Chen, and P. Zhang,
Solving the Sampling Problem of the Sycamore Quantum Circuits,
Phys. Rev. Lett. {\bf129}, 090502 (2022).

\bibitem {Aaronson2020}
S. Aaronson and S. Gunn,
On the Classical Hardness of Spoofing Linear Cross-Entropy Benchmarking,
THEORY OF COMPUTING {\bf16}, 1 (2020).

\bibitem {Gao2021}
X. Gao,et al., 
Limitations of linear cross-entropy as a measure for quantum advantage,
arXiv:2112.01657.


%%%%%%%%%%%%%%%%%%%%%%%%%%%%%%%%%%%%

\bibitem {Cross2019}
A. W. Cross,et al., 
Validating quantum computers using randomized model circuits, 
Phys. Rev. A. {\bf100}, 032328 (2019).

\bibitem {Pelofske2022}
E. Pelofske, A. Bartschi, and S. Eidenbenz,
Quantum Volume in Practice: What Users Can Expect from NISQ Devices (2022);
arXiv:2203.03816v4.

\bibitem {Wendin2022}
G. Wendin, et al.
Benchmarking the performance of variational quantum eigensolvers (VQE) applied to the HCN molecule,
Bulletin of the American Physical Society, 2022.

\bibitem {Lolur2021}
P. Lolur, M. Rahm, M. Skogh, L. García-Álvarez, and G. Wendin,
Benchmarking the variational quantum eigensolver through simulation of the ground state energy of prebiotic molecules on high-performance computers,
AIP Conference Proceedings {\bf2362}, 030005 (2021).

\bibitem {Mooney2021} 
G. J. Mooney, G. A. L. White, C. D. Hill and L. C. L. Hollenberg,
Generation and verification of 27-qubit Greenberger-Horne-Zeilinger states in a superconducting
quantum computer, J. Phys. Commun. {\bf5}, 095004 (2021).


%%%%%%%%%%%%%%%%%%%%%%%%%%%%%%

\bibitem {Farhi2014}
E. Farhi, J. Goldstone, and S. Gutmann, A Quantum Approximate Optimization Algorithm (2014);
arXiv:1411.4028.

\bibitem {Farhi2016}
E. Farhi and A. W. Harrow, 
Quantum Supremacy through the Quantum Approximate Optimization Algorithm, arXiv:1602.07674v1.

\bibitem {Farhi2022}
E. Farhi, J. Goldstone, S. Gutmann, and L. Zhou,
The Quantum Approximate Optimization Algorithm and the Sherrington-Kirkpatrick Model at Infinite Size,
Quantum {\bf6}, 759 (2022); arXiv:1910.08187v4.

\bibitem {ZhouLukin2020}
L. Zhou, et al.,
Quantum Approximate Optimization Algorithm: Performance, Mechanism, and Implementation on Near-Term Devices, 
Phys. Rev X {\bf10}, 021067 (2020).

\bibitem {Morales2020}
M. E. S. Morales, J. D. Biamonte and Z. Zimbor\'as,
On the universality of the quantum approximate optimization algorithm,
Quantum Information Processing {\bf19}: 291 (2020).

\bibitem {ZhangB2022}
B. Zhang, A. Sone, and Q. Zhuang,
Quantum computational phase transition in combinatorial problems,
npj Quantum Information {\bf8}:87 (2022).

\bibitem {Harrigan2021}
M. P. Harrigan, et al., Quantum approximate optimization of non-planar  graph problems on a planar superconducting processor, 
Nature Phys. {\bf17}, 332 (2021).

\bibitem {Herrman2021}
R. Herrman, J. Ostrowski, T. S. Humble, and G. Siopsis,
Lower bounds on circuit depth of the quantum approximate optimization algorithm,
Quantum Information Processing {\bf20}:59 (2021).

\bibitem {Borle2021}
A. Borle, V. E. Elfving and S. J. Lomonaco,
Quantum approximate optimization for hard problems in linear algebra, SciPost Phys. Core {\bf4}, 031 (2021).

\bibitem {Dupont2022}
M. Dupont, et al.,
An entanglement perspective on the quantum approximate optimization algorithm,
Phys. Rev. A {\bf106}, 022423 (2022).

\bibitem {Chen2022}
Y. Chen,et al., 
How much entanglement do quantum optimization algorithms require?, 
arXiv:2205.12283 (2022).

\bibitem {Sreedhar2022}
R. Sreedhar, et al., 
The quantum approximate optimization
algorithm performance with low entanglement and high circuit
depth, arXiv.2207.03404 (2022).

\bibitem {Willsch2020} 
M. Willsch, et al.,
Benchmarking the Quantum Approximate Optimization Algorithm,
Quantum Information Processing {\bf19}:197 (2020).

\bibitem {Bengtsson2020}
A. Bengtsson, et al.,
Quantum approximate optimization of the exact-cover problem on a superconducting quantum processor,
Phys. Rev. Applied {\bf14}, 034010  (2020)

\bibitem {Fitzek2021}
D. Fitzek, T. Ghandriz, L. Laine, M. Granath, and A. F. Kockum, 
Applying quantum approximate optimization to the heterogeneous vehicle routing problem, (2021);
arXiv:2110.06799.

\bibitem {Lacroix2020}
N. Lacroix, et al.,
Improving the Performance of Deep Quantum Optimization Algorithms with Continuous Gate Sets,
PRX Quantum {\bf1}, 020304 (2020)

\bibitem {Weggemans2022}
Jordi R. Weggemans, et al.,
Solving correlation clustering with QAOA and a Rydberg qudit system: a full-stack approach,
Quantum {\bf6}, 687 (2022); arXiv:2106.11672v3.

\bibitem {Verdon2019}
G. Verdon, et al.,
Learning to learn with quantum neural networks via classical neural networks,
arXiv:1907.05415v1.

\bibitem {Moussa2022}
C. Moussa, H. Wang, T. B\"ack, and V. Dunjko,
Unsupervised strategies for identifying optimal parameters in Quantum Approximate Optimization Algorithm,
EPJ Quantum Technology {\bf9}:11 (2022). 

\bibitem {Patel2022}
Y. J. Patel, S. Jerbi, T. B\"ack, and V. Dunjko,
Reinforcement Learning Assisted Recursive QAOA, arXiv:2207.06294v1.

\bibitem {Bravyi2020}
S. Bravyi, A. Kliesch, R. Koenig, and E. Tan
Obstacles to Variational Quantum Optimization from Symmetry Protection,
Phys. Rev. Lett. {\bf125}, 260505 (2020).

\bibitem {Egger2021}
D. J. Egger, J. Marecek, and S. Woerner,
Warm-starting quantum optimization,
Quantum  {\bf5}, 479 (2021); arXiv:2009.10095v4.

\bibitem {Guerreschi 2019} 
G. G. Guerreschi and A. Y. Matsuura,
QAOA for Max-Cut requires hundreds of qubits for quantum speed-up,
Scientific Reports {\bf9}: 6903 (2019).

\bibitem {Wurtz2022}
J. Wurtz and P. J. Love,
Counterdiabaticity and the quantum approximate optimization algorithm,
Quantum  {\bf6}, 635 (2022).

\bibitem {Chandarana2022}
P. Chandarana, et al.,
Digitized-counterdiabatic quantum approximate optimization algorithm,
Phys. Rev. Research  {\bf4}, 013141 (2022).

\bibitem {Hegade2022}
N. N. Hegade, X. Chen, and E. Solano,
Digitized counterdiabatic quantum optimization,
Phys. Rev. Research  {\bf4}, L042030 (2022).



%%%%%%%%%%%%%%%%%%%%%%
%%%%%%%%%%%%%%%%%%%%%%%

\bibitem {OngTeo2021}
J. H. Ong, and J. Teo,
Systematic Review and Open Challenges in Hyper-heuristics Usage On Expensive Optimization Problems with Limited Number of Evaluations,
2021 IEEE Symposium on Industrial Electronics \& Applications (ISIEA); DOI: 10.1109/ISIEA51897.2021.9509993

\bibitem {Monden2012}
Y. Monden, TOYOTA Production System An Integrated Approach to Just-In-Time, CRC Press, Taylor \& Francis Group, 2012.

\bibitem {deSousa2019}
W. Trigueiro de Sousa Junior, et al.,
Discrete simulation-based optimization methods for industrial engineering problems: A systematic literature review,
Computers \& Industrial Engineering 128, 526 (2019).

\bibitem {Csalodi2021}
R. Csalodi, et al.,
Industry 4.0-Driven Development of Optimization Algorithms: A Systematic Overview,
Complexity {\bf2021}:6621235 (2021).

\bibitem {Rai2021}
R. Rai, M. Kumar Tiwari, D. Ivanov and A. Dolgui,
Machine learning in manufacturing and industry 4.0 applications, 
International Journal of Production Research,  {\bf59} 4773 (2021).

\bibitem {Wedelin1995}
D. Wedelin, 
An algorithm for large scale 0-1 integer programming with application to airline crew scheduling,
Annals of Operations Research {\bf57}, 283 (1995).

\bibitem {Gronkvist2005}
M. Gr\"onkvist, 
The Tail Assignment Problem,
PhD thesis, 2005, Chalmers University of Technology.
 
 \bibitem {Vikstal2020} 
P. Vikstål, et al.,
Applying the Quantum Approximate Optimization Algorithm to the Tail-Assignment Problem,
Phys. Rev. Applied  {\bf14}, 034009 (2020),

\bibitem {Svensson2021}
M. Svensson, et al.,
A Hybrid Quantum-Classical Heuristic to solve large-scale Integer Linear Programs, arXiv:2103.15433v2.

\bibitem {Glover2022a}
F. Glover, G. Kochenberger, and Y. Du,
Quantum Bridge Analytics I: A Tutorial on Formulating and Using QUBO Models,  
Annuals of Operations Research  {\bf17}, 335?371 (2022).
 
\bibitem {Glover2022b}
F. Glover, G. Kochenberger, M. Ma, and Y. Du,
Quantum Bridge Analytics II: QUBO-Plus, network optimization and combinatorial chaining for asset exchange, 
Annals of Operations Research  {\bf314}, 185 (2022).

\bibitem {Ajagekar2022}
A. Ajagekar , K. Al Hamoud, and F. You,
Hybrid Classical-Quantum Optimization Techniques for Solving Mixed-Integer Programming Problems in Production Scheduling,
IEEE Transactions on Quantum Engineering  {\bf3}: 3102216 (2022). 

\bibitem {Willsch2022a}
D. Willsch, et al.,
Benchmarking Advantage and D-Wave 2000Q quantum annealers with exact cover problems,
Quantum Information Processing  {\bf21}:141 (2022).


%%%%%%%%%%%%%%%%%%%%%%%%%%%%%%


\bibitem {Lee2019} 
J. Lee, W. J. Huggins, M. Head-Gordon, and K. B. Whaley,
Generalized Unitary Coupled Cluster Wave functions for Quantum Computation,
J. Chem. Theory Comput. {\bf15}, 311 (2019).

\bibitem {Cao2019}
Y. Cao, et al.,
Quantum Chemistry in the Age of Quantum Computing,
Chem. Rev.  {\bf119}, 10856 (2019).

\bibitem {McArdle2020}
S. McArdle, et al., Quantum computational chemistry,
Rev. Mod. Phys. {\bf92}, 015003  (2020).

\bibitem {McClean2021}
J. R. McClean, et al.,
What the foundations of quantum computer science teach us about chemistry,
J. Chem. Phys. {\bf155}, 150901 (2021).

\bibitem {Fedorov2022} 
D. A. Fedorov, B. Peng, N. Govind, and Y. Alexeev,
VQE method: a short survey and recent developments,
Materials Theory {\bf6}:2 (2022).

\bibitem {Sokolov2020}
I. O. Sokolov, et al., Quantum orbital-optimized unitary coupled cluster methods in the strongly correlated regime: Can quantum algorithms outperform their classical equivalents?
J. Chem. Phys. {\bf152}, 124107 (2020).

\bibitem {Sugisaki2019}
K. Sugisaki, et al.,
Quantum Chemistry on Quantum Computers: A Method for Preparation of Multiconfigurational Wave Functions on Quantum Computers without Performing Post-Hartree-Fock Calculations, 
ACS Cent. Sci. {\bf5}, 167 (2019).

\bibitem {Sugisaki2022} 
K. Sugisaki, et al.,
Variational quantum eigensolver simulations with the multireference unitary coupled cluster ansatz: a case study of the C2v quasi-reaction pathway of beryllium insertion into a H 2 molecule,
Phys. Chem. Chem. Phys. {\bf24}, 8439 (2022).

\bibitem {Elfving2021}
V. E. Elfving, M. Millaruelo, J. A. G\'amez, and C. Gogolin,
Simulating quantum chemistry in the seniority-zero space on qubit-based quantum computers,
Phys. Rev. A {\bf1103}, 032605 (2021).

\bibitem {Sapova2022} 
M. D. Sapova, and A. K. Fedorov,
Variational quantum eigensolver techniques for simulating carbon monoxide oxidation
Communications Physics  {\bf5}:19 (2022).

\bibitem {Lolur2023}
Phalgun Lolur, et al.,
Reference-State Error Mitigation: A Strategy for High Accuracy Quantum Computation of Chemistry.
Journal of Chemical Theory and Computation (Jan 27, 2023); https://doi.org/10.1021/acs.jctc.2c00807.

\bibitem {ChenBooth2021}
H. Chen, M. Nusspickel, J. Tilly, and G. H. Booth,
Variational quantum eigensolver for dynamic correlation functions,
Phys. Rev. A {\bf104}, 032405 (2021).

\bibitem {YuZhang2022} 
Y. Zhang, et al.,
Variational quantum eigensolver with reduced circuit complexity,
npj Quantum Information {\bf8}:9 (2022).

\bibitem {JAGT2023}
J.A. de Gracia-Trevi\~no, M. G. Delcey, and G. Wendin,
Complete active space methods for NISQ devices: The importance of canonical orbital optimization for accuracy and noise resilience,
ChemRxiv (2023).

\bibitem {Grimsley2019}
H. R. Grimsley, S. E. Economou, E. Barnes, and N. J. Mayhall,
An adaptive variational algorithm for exact molecular simulations on a quantum computer,
Nature Commnunications {\bf10}:3007 (2019).

\bibitem {Tang2020}
H. L. Tang, et. al,
qubit-ADAPT-VQE: An adaptive algorithm for constructing hardware-efficient ans\"atze on a quantum processor,
PRX Quantum {\bf2}, 020310 (2021).

\bibitem {Tkachenko2021}
N. V. Tkachenko, et al.,
Correlation-Informed Permutation of Qubits for Reducing Ansatz Depth in the Variational Quantum Eigensolver,
PRX Quantum {\bf 2}, 020337 (2021).

\bibitem {Lan2022}
Z. Lan and W. Z. Liang,
Amplitude Reordering Accelerates the Adaptive Variational Quantum Eigensolver Algorithms,
J. Chem. Theory Comput.  {\bf18}, 5267 (2022).

\bibitem {LolurHPC2021}
The main results of \cite {Lolur2021} are produced with an iMac i9, 8-cores workstation with 128 GB RAM (max 32 qubits). The practical limit for the iMac to achieve an accurate ground-state energy minimum is so far connected with  H$_2$O (6-31G): 20 qubits and 73 000 gates, taking 66.5 hours to converge to 10$^{-6}$ Ha precision of  the ground-state energy for the equilibrium geometry. 
In the case of NCCN (STO-6G: 29 qubits, 2815 variational parameters, 515611 gates)  it took 132 CPU-hours (5.5 days) to produce the first energy value. This must then be multiplied by 2815 to evaluate one gradient vector for the first energy estimate by the optimizer. And then this will need to be iterated at least 10 times to converge to an accurate ground state energy value. The time-to-solution (TTS) is then about $3.7 \cdot 10^6$ CPU-hours. For comparison, SUMMIT at ORNL, USA, with in total 202752 cores, would finish the NCCN-instance in 147 hours = 6 days. In the QChem case, the challenge lies in the number of variational parameters and the number of gates needed, not in the number of qubits. Smart compilation can greatly reduce the number of gates \cite{Wendin2022}, but does not really change the poor scaling.
 
\bibitem {Calvin2021}
Justus A. Calvin, et al.,
Many-Body Quantum Chemistry on Massively Parallel Computers,
Chem. Rev.  {\bf121}, 1203 (2021).

\bibitem {Jones2022}
G. M. Jones, P. D. V. S. Pathirage, and K. D. Vogiatzis, 
Data-driven Acceleration of Coupled-Cluster Theory and Perturbation Theory Methods, in: "Quantum Chemistry in the Age of Machine Learning", 
2022, Editor: Pavlo Dral, Elsevier, pp. 509-529.

\bibitem {Scholl2021}
Pascal Scholl, et al.,
Quantum simulation of 2D antiferromagnets with hundreds of Rydberg atoms,
Nature  {\bf595}, 233 (2021).

\bibitem {Lamata2018}
L. Lamata, A. Parra-Rodriguez, M. Sanz, and E. Solano, 
Digital-analog quantum simulations with superconducting circuits,
Adv. Phys. X {\bf3}, 1457981 (2018).

\bibitem {Yu2022} 
Jing Yu, et al.,
Superconducting circuit architecture for digital-analog quantum computing,
EPJ Quantum Technology {\bf9}:9 (2022).

\bibitem {Karamlou2022}
A. H. Karamlou, et al.
Quantum transport and localization in 1d and 2d tight-binding lattices,
npj Quantum Information {\bf8}:35 (2022).

\bibitem {Landsman2019}
K. A. Landsman, et al., 
Verified quantum information scrambling,
Nature {\bf567}, 61 (2019). 

\bibitem {Touil2020}
Akram Touil and Sebastian Deffner,
Quantum scrambling and the growth of mutual information,
Quantum Sci. Technol.  {\bf5} 035005 (2020).

\bibitem {Mi2021}
X. Mi, Information scrambling in quantum circuits,
Sience  {\bf374}, 1479 (2021).

\bibitem {Braumuller2021}
J. Braum\"uller, et al.,
Probing quantum information propagation with out-of-time-ordered correlators,
Nature Physics  {\bf18}, 172 (2022).



%%%%%%%%%%%%%%%%%%%%%%%%%%%%%%%%%%%
%%%%%%%%%%%%%%%%%%%%%%%%%%%%%%%%%%%
%%%%%%%%%%%%%%%%%%%%%%%%%%%%%%%%%%%
%%%%%%%%%%%%%%%%%%%%%%%%%%%%%%%%%%%

\bibitem {CAPD}
Central Auditory Processing Disorder (CAPD) is a deficiency in the central auditory nervous system (CANS).
https://www.asha.org/practice-portal/clinical-topics/central-auditory-processing-disorder/
https://www.additudemag.com/central-auditory-processing-disorder/

\bibitem {Souffi2023}
S. Souffi, et al.,
Reduction in sound discrimination in noise is related to envelope similarity and not to a decrease in envelope tracking abilities, 
J. Physiol.  {\bf601.1}, 123 (2023).

\bibitem {CPS}
The "Cocktail Party Syndrome" is the name given to the inability to differentiate sound from background noise. 
https://en.wikipedia.org/wiki/Cocktail\_party\_effect\

\bibitem {Hashim2021}
 A. Hashim, et al., Randomized Compiling for Scalable Quantum Computing on a Noisy Superconducting Quantum Processor, Phys. Rev. X {\bf11}, 041039 (2021).

\bibitem {Ahsan2022}
M. Ahsan, S. Abbas Z. Naqvi, and H. Anwer
Quantum circuit engineering for correcting coherent noise,
Phys. Rev. A {\bf105}, 022428 (2022).

\bibitem {Vepsalainen2020}
A. P. Vepsäläinen, et al.,
Impact of ionizing radiation on superconducting qubit coherence,
Nature {\bf584}, 551 (2020).

\bibitem {Wilen2021}
C. D. Wilen, et al., 
Correlated charge noise and relaxation errors in superconducting qubits,
Nature {\bf594}, 369 (2021).

\bibitem {Burnett2019}
J. J. Burnett, et al, 
Decoherence benchmarking of superconducting qubits,
npj Quantum Information {\bf5}: 54 (2019).

\bibitem {Muller2019}
C. M\"uller, J. H. Cole and J. Lisenfeld,
Towards understanding two-level-systems in amorphous solids: insights from quantum circuits,
Rep. Prog. Phys. {\bf82}, 124501 (2019).

\bibitem {Bilmes2020}
A. Bilmes, et al.,
Resolving the positions of defects in superconducting quantum bits,
Sci. Rep. {\bf10}, 3090 (2020).

\bibitem {Caroll2022}
M. Carroll, et al., 
Dynamics of superconducting qubit relaxation times, npj Quantum Information {\bf8}:132 (2022).

\bibitem {Spring2022}
P. A. Spring, et al., 
High coherence and low cross-talk in a tileable 3D integrated superconducting circuit architecture, 
Sci. Adv. 8, eabl6698 (2022). 

\bibitem {PZhao2022}
P. Zhao, et al.,
Quantum Crosstalk Analysis for Simultaneous Gate Operations on Superconducting Qubits,
PRX Quantum {\bf3}, 020301 (2022).

\bibitem {Wittler2021}
N. Wittler, et al.,
Integrated Tool Set for Control, Calibration, and Characterization of Quantum Devices Applied to Superconducting Qubits,
Phys. Rev. Applied {\bf15}, 034080 (2021).  

\bibitem {Temme2017}
K. Temme, S. Bravyi, J. M. Gambetta, Error mitigation for short-depth quantum circuits.
Phys. Rev. Lett. {\bf119}, 180509 (2017).

\bibitem {YLi2017}
Y. Li and S. C. Benjamin, 
Efficient Variational Quantum Simulator Incorporating Active Error Minimization,
Phys. Rev. X {\bf7}, 021050 (2017).

\bibitem {Endo2018}
S. Endo, S. C. Benjamin, Y. and Li, 
Practical quantum error mitigation for near-future applications,
Physical Review X {\bf8}, 031027 (2018).

\bibitem {Song2019}
C. Song, et al.,
Quantum computation with universal error mitigation on a superconducting quantum processor,
Sci. Adv. 5: eaaw5686 (2019).

\bibitem {vandenBerg2022}
E. van den Berg, Z. K. Minev, A. Kandala, and K. Temme,
Probabilistic error cancellation with sparse Pauli-Lindblad models on noisy quantumvprocessors, 
arXiv:2201.09866v2.

\bibitem {Takagi2022}
R. Takagi, S. Endo, S. Minagawa, and M. Gu,
Fundamental limits of quantum error mitigation, 
npj Quantum Information {\bf8}:114 (2022).

\bibitem {Giurgica-Tiron2020}
T. Giurgica-Tiron, et al.,
Digital zero noise extrapolation for quantum error mitigation,
2020 IEEE International Conference on Quantum Computing and Engineering (QCE); DOI: 10.1109/QCE49297.2020.00045

\bibitem {Schultz2022}
K. Schultz, et al.,
Impact of time-correlated noise on zero-noise extrapolation,
Phys. Rev. A  {\bf106}, 052406 (2022).

\bibitem {Pascuzzi2022}
V. R. Pascuzzi, et al.,
Computationally efficient zero-noise extrapolation for quantum-gate-error mitigation,
Phys. Rev. A {\bf105}, 042406 (2022).

\bibitem {Leyton-Ortega2023}
V. Leyton-Ortega,  S. Majumder, and R. C. Pooser,
Quantum error mitigation by hidden inverses protocol in superconducting quantum devices,
Quantum Sci. Technol. {\bf8}, 014008 (2023).


%%%%%%%%%%%%%%%%%%%%%%%%%%%%%%%%%%%
%%%%%%%%%%%%%%%%%%%%%%%%%%%%%%%%%%%

\bibitem {Fowler2012}
A. G. Fowler, et al. 
Surface codes: towards practical large-scale quantum computation,
Phys Rev A. {\bf86}, 032324 (2012).

\bibitem {Terhal2015}
B. M. Terhal,
Quantum error correction for quantum memories,
Rev. Mod. Phys {\bf87}, 307 (2015).

\bibitem {Roffe2019}
J. Roffe, 
Quantum error correction: an introductory guide,
Contemporary Physics, {\bf60}:3, 226 (2019).
 
 \bibitem {Ryan-Anderson2021}
C. Ryan-Anderson, et al.,
Realization of Real-Time Fault-Tolerant Quantum Error Correction,
Phys. Rev, X {\bf11}, 041058 (2021)

\bibitem {Ryan-Anderson2022}
C. Ryan-Anderson, et al.,
 Implementing Fault-tolerant Entangling Gates on the Five-qubit Code and the Color Code,
 arXiv: 2208.01863v1
 
\bibitem {Postler2022}
L. Postler, et al.,
Demonstration of fault-tolerant universal quantum gate operations,
Nature {\bf605}, 675 (2022).

\bibitem {Kelly2015}
J. Kelly. et al.,
State preservation by repetitive error detection in a superconducting quantum circuit,
Nature {\bf519}, 66 (2015).

\bibitem {Ouyang2021}
Y. Ouyang,
Avoiding coherent errors with rotated concatenated stabilizer codes,
npj Quantum Information {\bf7}:87 (2021).

\bibitem {Piveteau2021}
C Piveteau, et al.,
Error Mitigation for Universal Gates on Encoded Qubits,
Phys. Rev. Lett. {\bf127}, 200505 (2021).

\bibitem {Suzuki2022}
Y. Suzuki, S. Endo, K. Fujii, and Y. Tokunaga,
Quantum Error Mitigation as a Universal Error Reduction Technique: Applications from the NISQ to the Fault-Tolerant Quantum Computing Eras, 
PRX Quantum {\bf3}, 010345 (2022). 



%%%%%%%%%%%%%%%%%%%%%%%%%%%%%%%%%%%
%%%%%%%%%%%%%%%%%%%%%%%%%%%%%%%%%%%


\bibitem {Osprey2022}
https://newsroom.ibm.com/2022-11-09-IBM-Unveils-400-Qubit-Plus-Quantum-Processor-and-Next-Generation-IBM-Quantum-System-Two 

\bibitem {Piveteau2022}
C. Piveteau, and D. Sutter, Circuit knitting with classical communication; arXiv:2205.00016.

\bibitem {Garcia-Perez2020}
G. Garc\'ia-P\'erez, M. A. C. Rossi, and S. Maniscalco,
IBM Q Experience as a versatile experimental testbed for simulating open quantum systems,
npj Quantum Information {\bf6}:1 (2020).

\bibitem {Wack2021}
A. Wack, et al.,
Scale, Quality, and Speed: three key attributes to measure the performance of near-term quantum computers,
arXiv:2110.14108v2

\bibitem {Gambetta2022b}
J. Gambetta,
IBM Quantum Summit 2022,
IBM Quantum State of the Union 2022,
https://www.ibm.com/quantum/summit 

%%%%%%%%%%%%%%%%%%%%%%%%%%%%%%%%%%%
%%%%%%%%%%%%%%%%%%%%%%%%%%%%%%%%%%%


\bibitem {Chamberland2022}
C. Chamberland, et al.,
Building a Fault-Tolerant Quantum Computer Using Concatenated Cat Codes,
PRX Quantum {\bf3}, 010329 (2022).

\bibitem {Karzig2017}
T. Karzig, et al.,
Scalable designs for quasiparticle-poisoning-protected topological quantum computation with majorana zero modes,
Phys. Rev. B {\bf95}, 235305 (2017).



%%%%%%%%%%%%%%%%%%%%%%%%%%%%%%%%%%%
%%%%%%%%%%%%%%%%%%%%%%%%%%%%%%%%%%%
%%%%%%%%%%%%%%%%%%%%%%%%%%%%%%%%%%%
%%%%%%%%%%%%%%%%%%%%%%%%%%%%%%%%%%%
%%%%%%%%%%%%%%%%%%%%%%%%%%%%%%%%%%%
%%%%%%%%%%%%%%%%%%%%%%%%%%%%%%%%%%%
%%%%%%%%%%%%%%%%%%%%%%%%%%%%%%%%%%%
%%%%%%%%%%%%%%%%%%%%%%%%%%%%%%%%%%%




\bibitem {Reiher2017}
M. Reiher, et al.,
Elucidating reaction mechanisms on quantum computers,
PNAS {\bf114}, 7555 (2017).

\bibitem {Berry2019} 
D. W. Berry, et al.,
Qubitization of Arbitrary Basis Quantum Chemistry by Low Rank Factorization,
Quantum {\bf3}, 208 (2019); arXiv:1902.02134v4.

\bibitem {Liu2022} 
H. Liu, et al.,
Prospects of quantum computing for molecular sciences, 
Materials Theory {\bf6}:11 (2022).

\bibitem {Lee2022} 
S. Lee, et al.,
Is there evidence for exponential quantum advantage in quantum chemistry?;
arXiv:2208.02199v2.

\bibitem {Wang2021}
S. Wang, et al.,
Noise-induced barren plateaus in variational quantum algorithms,
Nature Communications {\bf12}:6961 (2021).

\bibitem {Bittel2021}
 L. Bittel and M. Kliesch, 
 Training variational quantum algorithms is NP-hard, 
 Phys. Rev. Lett. {\bf127}, 120502 (2021).

\bibitem {EwinTang2022}
E. Tang, 
Dequantizing algorithms to understand quantum advantage in machine learning,
Nature Reviews Physics {\bf4}, 693 (2022).

\bibitem {EwinTang2021}
E. Tang,
Quantum Principal Component Analysis Only Achieves an Exponential Speedup Because of Its State Preparation Assumptions,
Phys. Rev. Lett. {\bf127}, 060503 (2021).

\bibitem {Cotler2021}
J. Cotler, H.-Y. Huang, and J. R. McClean,
Revisiting dequantization and quantum advantage in learning tasks,
arXiv:2112.00811v2.
 
\bibitem {Gharibian2022} 
S. Gharibian and F. Le Gall,
Dequantizing the Quantum Singular Value Transformation: Hardness and Applications to Quantum Chemistry and the Quantum PCP Conjecture,
STOC 2022: Proceedings of the 54th Annual ACM SIGACT Symposium on Theory of ComputingJune (STOC 2022), pp.19-32;  arXiv:2111.09079v4.

\bibitem {Lloyd2016}
S. Lloyd, S. Garnerone, and P. Zanardi,
Quantum algorithms for topological and geometric analysis of data,
Nature Communications {\bf6}:10138 (2016).

\bibitem {Berry2022}
D. W. Berry, et al.,
Quantifying Quantum Advantage in Topological Data Analysis,
arXiv:2209.13581v1.

\bibitem {Akhalwaya2022} 
I. Y. Akhalwaya, et al.,
Towards Quantum Advantage on Noisy Quantum Computers;
arXiv:2209.09371v2.

\bibitem {Huang2022}
H.Y. Huang et al., Quantum advantage in learning from experiments, 
Science {\bf376}, 1182 (2022). 

\bibitem {Sullivan2022}
J. O'Sullivan, et al.,
Random-Access Quantum Memory Using Chirped Pulse Phase Encoding,
PRX {\bf12}, 041014 (2022). 


%%%%%%%%%%%%%%%%%%%%%%%%%%%%%%%%
%%%%%%%%%%%%%%%%%%%%%%%%%%%%%%%%

\bibitem {Calzona2023}
A. Calzona and M. Carrega,
Muti-mode architectures for noise-resilient superconducting qubits,
Supercond. Sci. Technol. {\bf36}, 023001 (2023).

\bibitem {Place2021}
A. P. M. Place, et al.,
New material platform for superconducting transmon qubits with coherence times exceeding 0.3 milliseconds,
Nature Communications {\bf12},1779 (2021).

\bibitem {CWang2022}
C. Wang, et al.,
Towards practical quantum computers: transmon qubit with a lifetime approaching 0.5 milliseconds,
npj Quantum Information {\bf8}:3 (2022).

\bibitem {Tennant2022}
D. M. Tennant, et al.,
Low-Frequency Correlated Charge-Noise Measurements Across Multiple Energy Transitions in a Tantalum Transmon, 
PRX Quantum {\bf3}, 030307 (2022).

\bibitem {WendinShumeiko2021}
G. Wendin and V. S. Shumeiko, 
Coherent manipulation of a spin qubit,
Science  {\bf373}, 390 (2021),

\bibitem {DasSarma2015}
S. Das Sarma, M. H. Freedman, and C. Nayak, 
Majorana zero modes and topological quantum computation,
npj Quantum Inf. {\bf1}, 15001 (2015).

\bibitem {Pikulin2021}
Dmitry I. Pikulin,et al.,
Protocol to identify a topological superconducting phase in a three-terminal device,
arXiv:2103.12217v1.

\bibitem {Aghaee2022}
 M. Aghaee et al.,
 InAs-Al Hybrid Devices Passing the Topological Gap Protocol,
 arXiv:2207.02472v3.



%%%%%%%%%%%%%%%%%%%%%%%
%%%%%%%%%%%%%%%%%%%%%%%

\bibitem {Zhong2021}
Y. Zhong, et al., 
Deterministic multi-qubit entanglement in a quantum network,
Nature {\bf590}, 571 (2021).

\bibitem {Yan2022}
H. Yan, et al., 
Purification and Protection in a Superconducting Quantum Network
Phys. Rev. Lett. {\bf128}, 080504 (2022).

\bibitem {Chu2020}
Y. Chu, and S. Gr\"oblacher, 
A perspective on hybrid quantum opto- and electromechanical systems,
Appl. Phys. Lett. {\bf117}, 150503 (2020). 

\bibitem {Arnold2020}
G. Arnold, et al.,
Converting microwave and telecom photons with a silicon photonic nanomechanical interface, 
Nat. Commun. {\bf11}, 4460 (2020).

 \bibitem {Wang2022}
C. Wang, et al.,
High-efficiency microwave-optical quantum transduction based on a cavity electro-optic superconducting system with long coherence time, 
npj Quantum Information npj Quantum Information {\bf8}:149 (2022).

\bibitem {Kumar2022}
A. Kumar, et al.,
Quantum-limited millimeter wave to optical transduction,
arXiv:2207.10121v1.

\bibitem {DiAdamo2021} 
S. DiAdamo, M. Ghibaudi, and J. Cruise,
Distributed Quantum Computing and Network Control for Accelerated VQE,
IEEE Transactions on Quantum Engineering {\bf2}: 3100921 (2021).




%%%%%%%%%%%%%%%%%%%%%%%
%%%%%%%%%%%%%%%%%%%%%%%

\bibitem {Sivak2022b}
V. V. Sivak, et al.,
Real-time quantum error correction beyond break-even,
arXiv:2211.09116v1.

\bibitem {Sivak2022a}
V. V. Sivak, et al.,
Model-Free Quantum Control with Reinforcement Learning,
Phys. Rev. X {\bf12}, 011059 (2022).

\bibitem {Kudra2022}
M. Kudra, et al.,
Experimental realization of deterministic and selective photon addition in a bosonic mode assisted by an ancillary qubit, arXiv:2212.12079v1.

\bibitem {Guillaud2019}
J. Guillaud and M. Mirrahimi,
Repetition cat qubits for fault-tolerant quantum computation,
Phys. Rev. X {\bf9}, 041053 (2019).

\bibitem {Axline2018}
 C. J. Axline, et al.,
On-demand quantum state transfer and entanglement between remote microwave cavity memories,
Nature Physics {\bf14}, 705 (2018).

\bibitem {Chu2017}
Y. Chu, et al.,
Quantum acoustics with superconducting qubits, 
Science {\bf358}, 199 (2017).

\bibitem {vonLupke2022}
Uwe von L\"upke, et al.,
Parity measurement in the strong dispersive regime of circuit quantum acoustodynamics,
Nature Physics {\bf18}, 794 (2022). 

\bibitem {Schrinski2022}
B. Schrinski, et al.,
Macroscopic quantum test with bulk acoustic wave resonators,
arXiv:2209.06635v1.

\bibitem {Wollack2022}
E. Alex Wollack, Agnetta Y. Cleland, Rachel G. Gruenke, Zhaoyou Wang, Patricio Arrangoiz-Arriola, and Amir H. Safavi-Naeini,
Quantum state preparation and tomography of entangled mechanical resonators,
Nature  {\bf604}, 463 (2022).


 
%%%%%%%%%%%%%%%%%%%%%%%
%%%%%%%%%%%%%%%%%%%%%%%
%%%%%%%%%%%%%%%%%%%%%%%
%%%%%%%%%%%%%%%%%%%%%%%




\bibitem {Fish2021}
J. Fish, G. J. Wagner, and S. Keten,
Mesoscopic and multiscale modelling in materials,
Nature Materials {\bf20}, 774 (2021).

\bibitem {Warshel2014}
A. Warshel,
Multiscale Modeling of Biological Functions: From Enzymes to Molecular Machines (Nobel Lecture),
Angew. Chem. Int. Ed. {\bf53}, 10020 (2014).

\bibitem {Cheng2020}
H.-P. Cheng, et al.,
Application of Quantum Computing to Biochemical Systems: A Look to the Future, 
Front. Chem. {\bf8}:587143 (2020). 

\bibitem {Barabasi2011}
A. L. Barabasi, N. Gulbahce, and J. Loscalzo, 
Network medicine: a network-based approach to human disease,
Nat. Rev. Genet. {\bf12}:56e68 (2011).

\bibitem {LeeLoscalzo2019}
L. Y.-H. Lee and J. Loscalzo,
Network Medicine in Pathobiology,
Am. J. Pathol. {\bf189}, 1311 (2019).

\bibitem {Santos2021} 
S. de Siqueira Santos, et al.,
Machine learning and network medicine approaches for drug repositioning for COVID-19,
Patterns (N Y) {\bf3}(1):100396 (2022); doi: 10.1016/j.patter.2021.100396.

\bibitem {Infante2021}
T. Infante, et al.,
Machine learning and network medicine: a novel approach for precision medicine and personalized therapy in cardiomyopathies.
Journal of Cardiovascular Medicine {\bf22}(6), 429 (2021).

\bibitem {Maniscalco2022}
S. Maniscalco, et al.,
Quantum network medicine: rethinking medicine with network science and quantum algorithms,
arXiv:2206.12405v1.

\bibitem {Goings2022}
J. J. Goings, et al.,
Reliably assessing the electronic structure of cytochrome P450 on today's classical computers and tomorrow's quantum computers,
PNAS  {\bf119}: e2203533119 (2022).

\bibitem {Wendin2028} 
G. Wendin, Five years later, what happened?
In preparation.

%%%%%%%%%%%%%%%%%%%%%%%%%%%%%%%%%%
%%%%%%%%%%%%%%%%%%%%%%%%%%%%%%%%%%
%%%%%%%%%%%%%%%%%%%%%%%%%%%%%%%%%%
%%%%%%%%%%%%%%%%%%%%%%%%%%%%%%%%%%
%%%%%%%%%%%%%%%%%%%%%%%%%%%%%%%%%%
%%%%%%%%%%%%%%%%%%%%%%%%%%%%%%%%%%

 
\end{thebibliography}
\end{document}